\documentclass[prd,twocolumn,floatfix,nopacs,preprintnumbers,nofootinbib]{revtex4}
\usepackage[utf8x]{inputenc}
\usepackage{mathrsfs}
\usepackage{amssymb}
\usepackage{amsmath}
\usepackage{mathtools}
\usepackage{slashed}
\usepackage{graphicx}

\usepackage[caption = false]{subfig}
\usepackage[normalem]{ulem}
\usepackage{xcolor}
\definecolor{lcolor}{rgb}{0.5,0,0}
\definecolor{citcolor}{rgb}{0,0.3,0.0}

\newlength{\mycol}
\usepackage[breaklinks,colorlinks,urlcolor=blue,citecolor=citcolor,linkcolor=lcolor]{hyperref}
\usepackage{mciteplus}

\newcommand{\rt}{{\mathbf{r}}}

\newcommand{\bt}{{\mathbf{b}}}

\newcommand{\kt}{{\mathbf{k}}}
\newcommand{\qt}{{\mathbf{q}}}

\newcommand{\nc}{N_\mathrm{c}}
\newcommand{\mcnr}{m_{c,NR}}
\newcommand{\xpom}{{x_\mathbb{P}}}

\newcommand{\gev}{\ \textrm{GeV}}
\newcommand{\fm}{\ \textrm{fm}}

\newcommand{\eq}{Eq.~}

\newcommand{\eqs}{Eqs.~}

\newcommand{\as}{\alpha_\mathrm{s}}
\newcommand{\der}{\mathrm{d}}

\newcommand{\Deltat}{\mathbf{\Delta}}

\newcommand{\jpsi}{$\mathrm{J}/\psi$ }
\newcommand{\jpsim}{\mathrm{J}/\psi}

\begin{document}
\author{T. Lappi}
\author{H. Mäntysaari}
\author{J. Penttala}
\affiliation{
Department of Physics, University of Jyväskylä %
 P.O. Box 35, 40014 University of Jyväskylä, Finland
}
\affiliation{
Helsinki Institute of Physics, P.O. Box 64, 00014 University of Helsinki, Finland
}

\title{
Relativistic corrections to the vector meson light front wave function
}

\preprint{}

\begin{abstract}
We compute a light front wave function for heavy vector mesons based on long distance matrix elements constrained by decay width analyses in the Non Relativistic QCD framework. Our approach provides a systematic expansion of the wave function in quark velocity. The first relativistic correction included in our calculation is found to be significant, and crucial for a good description of the HERA exclusive \jpsi\ production data. When looking at cross section ratios between nuclear and proton targets, the wave function dependence does not cancel out exactly. In particular the fully non-relativistic limit is found not to be a reliable approximation even in this ratio. The important role of the Melosh rotation to express the rest frame wave function on the light front is illustrated.

\end{abstract}

\maketitle

\section{Introduction}

At large densities or small Bjorken-$x$, non-linear QCD dynamics is expected to manifest itself  in  nuclear structure. To describe the QCD matter in this non-linear regime, an effective field theory known as the Color Glass Condensate (CGC) has been developed, see e.g.~\cite{Gelis:2010nm,Albacete:2014fwa}. 
Diffractive scattering processes at high energies are especially powerful probes of this region of  phase space. 
The advantage in diffractive, with respect to inclusive, scattering is that since no color charge transfer is allowed, even at leading order in perturbative QCD at least two gluons have to be  exchanged with the target. Consequently, the cross sections approximatively probe the square of the  gluon density~\cite{Ryskin:1992ui}, and can be expected to be highly sensitive to non-linear dynamics.

An especially interesting diffractive process is exclusive vector meson production in collisions of real or virtual photons with the target, where only one  meson with the same quantum numbers as the photon is produced. In these processes only vacuum quantum numbers are exchanged between the target and the diffractive system. Thus the target can remain intact, and the transverse momentum transfer can be used to probe the spatial structure of the target. This momentum transfer is by definition  the Fourier conjugate to the impact parameter. As such, it becomes possible to study the target structure differentially in the transverse plane. A particularly important channel is the production of \jpsi mesons. The charm quark is heavy enough to enable a weak coupling description of its elementary interactions. Nevertheless the quark mass is not large enough to make the process insensitive to saturation effects. Also experimentally the \jpsi is relatively easily identifiable and produced with large enough cross sections to be seen.

Exclusive \jpsi production in electron-proton deep inelastic scattering has been studied in detail at HERA by the H1 and ZEUS experiments~\cite{Aktas:2005xu,Chekanov:2002xi,Chekanov:2002rm,Aktas:2003zi,Chekanov:2004mw,Alexa:2013xxa}. Additionally, lighter $\rho$ and $\phi$~\cite{Adloff:1999kg,Chekanov:2005cqa,Aaron:2009xp} and heavier $\Upsilon$ states~\cite{Chekanov:2009zz,Adloff:2000vm} have been measured. 
Recently, it has also become possible to measure exclusive vector meson production  at RHIC and at the LHC in ultra peripheral collisions~\cite{Bertulani:2005ru,Klein:2019qfb} where the impact parameter between the two hadrons is large enough such that the scattering is mediated by quasi-real photons, see Refs.~\cite{Afanasiev:2009hy,TheALICE:2014dwa,Acharya:2018jua,Abbas:2013oua,Abelev:2012ba,Khachatryan:2016qhq,Sirunyan:2018sav,Adam:2019rxb,LHCb:2018ofh,Li:2020ntd} for recent measurements. These developments have also enabled vector meson photoproduction studies with nuclear targets, which are more sensitive to gluon saturation. Indeed  signatures of strong nuclear effects (e.g. saturation, or gluon shadowing) are seen in \jpsi photoproduction (see e.g. Refs.~\cite{Lappi:2013am,Guzey:2013qza, Guzey:2016piu}). The effects seen in these exclusive processes are consistent with inclusive measurements such as particle spectra in proton-nucleus collisions (see e.g.~\cite{Albacete:2010bs,Albacete:2012xq,Lappi:2013zma,Ducloue:2015gfa,Ducloue:2016pqr,Mantysaari:2019nnt}). However, in exclusive scattering  the non-linear effects are larger, since inclusive  cross sections at leading order are only sensitive to the first power of the gluon density.  

One major source of model uncertainties in the theoretical description of vector meson production follows from the nonperturbative vector meson wave function. For the \jpsi\!, a natural first approximation is to treat it as a fully nonrelativistic bound charm-anticharm state, which is the limit taken in the seminal work in Ref.~\cite{Ryskin:1992ui}. The calculation of Ref.~\cite{Brodsky:1994kf} recovers the same nonrelativistic result in the dipole picture (see also Ref.~\cite{Anand:2018zle}). Already early on, it has been argued that this nonrelativistic approximation obtains important corrections from the motion of the charm quark pair in the bound state~\cite{Hoodbhoy:1996zg,Frankfurt:1997fj}. More recently, much of the phenomenological literature on \jpsi photoproduction has used phenomenological light cone wave functions to describe the meson bound state. This has the advantage that the light cone wave function is invariant under boosts in the longitudinal direction, and is thus naturally more suited to high energy collision phenomena. A disadvantage of some recent phenomenological parametrizations has been that they do not fully use the information on the nonperturbative bound state physics, most importantly decay widths, of quarkonium states that are usually analyzed in terms of nonrelativistic wavefunctions.

In recent literature, the applied different phenomenological wave functions result in e.g. \jpsi production cross sections that differ up to $\sim 30\%$ from each other~\cite{Mantysaari:2017dwh,Lappi:2013am,Lappi:2010dd}. This is a large model uncertainty, compared to the precise data that is already available from HERA and the LHC, and especially given that the Electron Ion Collider (EIC)~\cite{Accardi:2012qut,Aschenauer:2017jsk} is in the horizon (and similar plans exist at CERN~\cite{AbelleiraFernandez:2012cc} and in China~\cite{Chen:2018wyz}). The EIC will perform vast amounts of precise DIS measurements over a wide kinematical range, which calls for robust theoretical predictions.

To reduce the model uncertainty related to the vector meson wave function, we propose in this work a new method to constrain the wave function for heavy mesons based on input from the Non Relativistic QCD (NRQCD) matrix elements. These matrix elements capture non-perturbative long-distance physics, and can be obtained by computing the vector meson decay widths in different channels as a systematic expansion in both the coupling constant $\as$ and the quark velocity $v$. As we will demonstrate, these matrix elements can be used to determine the value and the derivative of the vector meson wave function at the origin. As such, this approach provides more constraints than the phenomenological parametrizations widely used in the literature. In particular, starting from manifestly rotationally invariant rest frame wave functions, one by construction obtains consistent parametrizations of longitudinally and tranversally polarized vector mesons simultaneously, which is not obvious in many light cone approaches.

This manuscript is organized as follows. First, in Sec.~\ref{sec:dipolepicture} we review how  vector meson production is computed in the dipole picture within the Color Glass Condensate framework, and how the cross section depends on the vector meson light front wave function.  In Sec.~\ref{sec:nrqcd} we first present how to obtain the rest frame wave function in terms of the NRQCD matrix elements, and then show how this is transformed to the light cone by applying the Melosh rotation~\cite{Melosh:1974cu,Hufner:2000jb}. We compare the obtained NRQCD based wave function to other widely used wave functions that are reviewed in Sec.~\ref{sec:pheno-wf}. The numerical analysis including vector meson-photon overlaps and \jpsi production cross sections is presented in Sec.~\ref{sec:numerics}.

\section{Vector meson production in the dipole picture}
\label{sec:dipolepicture}
\subsection{Exclusive scattering}
At high energies exclusive vector meson production in virtual photon-proton (or nucleus) scattering can be described in a factorized form. The necessary ingredients are the virtual photon wave function $\Psi_\gamma^\lambda$ describing the $\gamma^* \to q\bar q$ splitting, the dipole-target scattering amplitude $N$ and the vector meson wave function $\Psi_V$ describing the transition $q\bar q \to V$. The scattering amplitude reads~\cite{Kowalski:2006hc} (note that the correct phase factor coupling the dipole size $\rt$ to the transverse momentum transfer $\mathbf{\Delta}$ is determined in Ref.~\cite{Hatta:2017cte})
\begin{multline}
\label{eq:vm_amp}
    \mathcal{A}^\lambda = 2i\int \der^2 \bt \der^2 \rt \frac{\der z}{4\pi} e^{-i \left(\bt + \left(\frac{1}{2}-z\right)\rt\right) \cdot \mathbf{\Delta}} \\
    \times \Psi_\gamma^{\lambda *}(\rt,Q^2,z) \Psi_V(\rt,z) N(\rt,\bt,\xpom).
\end{multline}
Here $Q^2$ is the photon virtuality, $\rt$ the transverse size of the dipole, $\bt$ the impact parameter  and $z$ the fraction of the photon light cone plus  momentum carried by the quark. The photon polarization is $\lambda$, with $\lambda=\pm 1$ referring to the transverse polarization and $\lambda=0$ to the longitudinal one.

In this work will study coherent vector meson $V$ production. The coherent cross section refers to the scattering process where the target proton (or nucleus) remains intact. In this case, the cross section as a function of squared momentum transfer $t\approx -\Deltat^2$ can be written as
\begin{equation}
\label{eq:vm_xs}
    \frac{\der \sigma^{\gamma^*p \to Vp}}{\der t} = R_g^2\left(1+\beta^2\right)\frac{1}{16\pi} |\mathcal{A}_{T,L}|^2.
\end{equation}
The dipole amplitude $N$ depends on the longitudinal momentum fraction $\xpom$ the target loses in the scattering process, which reads
\begin{equation}
    \xpom = \frac{M_V^2+Q^2 - t}{W^2+Q^2 - m_N^2}.
\end{equation}
Here $M_V$ is the mass of the vector meson $V$ and $m_N$ is the proton mass. The scattering amplitude $\mathcal{A}_{T,L}$ is obtained from Eq.~\eqref{eq:vm_amp} by summing over the quark helicities, and in the case of transverse (T) polarization, averaging over the photon polarization states $\lambda = \pm 1$.

In Eq.~\eqref{eq:vm_xs} two phenomenological corrections are included following Ref.~\cite{Kowalski:2006hc}. First, $\beta = \tan \left( \frac{\pi \delta}{2} \right)$ is the ratio between the real and imaginary parts of the scattering amplitude. It can be obtained from an analyticity argument as
\begin{equation}
    \delta = \frac{\partial \ln \mathcal{A}_{T,L}}{\partial \ln (1/\xpom)}.
\end{equation}

The so called skewedness correction is included in terms of the factor $R_g$, which reads,
\begin{equation}
    R_g = \frac{2^{2\delta+3}}{\sqrt{\pi}} \frac{\Gamma(\delta+5/2)}{\Gamma(\delta+4)}.
\end{equation}
This correction can be derived by considering the vector meson production in the  two-gluon exchange limit, assuming that the two gluons carry very different fractions of the target longitudinal momentum~\cite{Shuvaev:1999ce}. In this case, the cross section can be related to the collinearly factorized parton distribution functions scaled by the factor $R_g$. In the dipole picture applied here, where the two quarks are color rotated in the target color field and undergo multiple scattering, this limit is not reached. In this work we include both of the real part and skewedness corrections widely used in the previous literature, but emphasize that these numerically large corrections should be used with caution when predicting absolute normalizations for the cross sections.

In addition to  coherent scattering, one can study  incoherent diffraction where the target breaks up, but there is still no exchange of color charge between the produced vector meson and the target remnants. These processes are recently studied extensively in the literature as they probe, in addition to saturation effects~\cite{Lappi:2010dd}, also the event-by-event fluctuations of the scattering amplitude resulting from the target structure fluctuations, see e.g. Refs.~\cite{Mantysaari:2016ykx,Cepila:2016uku,Mantysaari:2016jaz,Traini:2018hxd} or Ref.~\cite{Mantysaari:2020axf} for a review. As the focus in this work is on the vector meson wave function which enters in calculations of both incoherent and coherent cross sections similarly, from now on we only consider  coherent scattering here.

\subsection{Virtual photon wave function}
The virtual photon splitting to a $c\bar c$ dipole is a simple QED process, and the photon wave function $\Psi_\gamma$ can be computed directly by applying the light cone perturbation theory (see e.g.~\cite{Brodsky:1997de,Kovchegov:2012mbw}).  Using the diagrammatic rules of light front perturbation theory and the conventions used in Refs.~\cite{Lepage:1980fj,Dosch:1996ss}, the wave function can be written as
\begin{equation}
\label{eq:photon_wf_spinor}
    \Psi^\lambda_{\gamma,h\bar h}(k) = \frac{e_f e\sqrt{\nc}}{q^- - k^- - k'^-}  \frac{\bar u_h(k) \slashed{\varepsilon}^\lambda(q) v_{\bar h}(k')}{ 4\sqrt{\pi} k^+ k'^+ q^+}.
\end{equation}
Here $e_f$ is the fractional charge of the quark (in this work we consider only charm quarks with $e_f=2/3$), $k$, $k'$ and $q$ are the quark, antiquark and photon momenta respectively, $e=\sqrt{4\pi \alpha_\text{em}}$ and $h$ and $\bar h$ refer to the quark and antiquark light front helicities~\cite{Soper:1972xc}. The factor $\sqrt{\nc}$ is included to obtain a squared wave function proportional to the number of colors $\nc$. The spinors which are the eigenstates of light front helicity read, in the Lepage-Brodsky convention
\begin{align}
\label{eq:uh}
        u_h(k) &= \frac{1}{2^{1/4}\sqrt{p^+}} \left( \sqrt{2}p^+ + \gamma^0 m + \alpha_T \cdot \kt \right)   \bar \chi_h \\
        \label{eq:vh}
        v_h(k) &= \frac{1}{2^{1/4}\sqrt{p^+}} \left( \sqrt{2} p^+ - \gamma^0 m + \alpha_T \cdot \kt \right)   \bar \chi_{-h},
\end{align}
where the four-component helicity spinors read $\bar \chi_{h=+1}=\frac{1}{\sqrt{2}} (1,0,1,0)^T$ and $\bar \chi_{h=-1}=\frac{1}{\sqrt{2}} (0,1,0,-1)^T$, and $\alpha_T = (\gamma^0\gamma^1, \gamma^0\gamma^2)$. We use the light cone variables defined as $p^\pm = \frac{1}{\sqrt{2}}(p^0 \pm p^3)$. The spinor normalization convention is $\bar u_h u_{\bar h}  = -\bar v_h v_{\bar h} = 2m\delta_{h\bar h}$, where $m$ is the quark mass.

In the light cone gauge, in which $\varepsilon^+=0$, the photon polarization vectors read
\begin{align}
    \varepsilon^{\lambda=0}(q) &= \left(0,0,0,\frac{Q}{q^+}\right) \\
    \varepsilon^{\lambda=\pm 1} &= \left(0, \varepsilon^\lambda_T,\frac{\qt \cdot \varepsilon^\lambda_T}{q^+}\right), 
\end{align}
where
\begin{equation}
\label{eq:trans_pol}
    \varepsilon^{\lambda=\pm 1}_T = (\mp 1, - i)/\sqrt{2},
    \end{equation}
and $Q^2=-q^2$. 

The wave function can be evaluated by substituting the polarization vectors and explicit expressions for the spinors in Eq.~\eqref{eq:photon_wf_spinor} and setting the photon transverse momentum $\qt$ to zero. It is convenient here to define a wave function in terms of the momentum fraction $z$ and pull out a factor $4\pi$. This should be done so that probability is conserved,
\begin{equation}
    \int \der k^+ | \Psi^\lambda_{\gamma,h\bar h}(k^+, \kt) |^2 = \int \frac{\der z}{4\pi}  |\Psi^\lambda_{\gamma,h\bar h}(z,\kt)|^2 ,
\end{equation}
so that we can write $\Psi_{\gamma,h\bar h}^\lambda (z, \kt) = \sqrt{4\pi q^+} \Psi_{\gamma,h\bar h}^\lambda(k^+, \kt)$. In momentum space, the wave functions read
\begin{equation}
\label{eq:photon_k_L}
    \Psi^{\lambda=0}_{\gamma,h\bar h}(z,\kt) = -e_f e \sqrt{N_c} \frac{2 Q z(1-z)}{(\kt^2 + \epsilon^2) } \delta_{h,-\bar h} 
\end{equation}
\begin{align}
\label{eq:photon_k_T}
    \Psi^{\lambda=+1}_{\gamma,h\bar h}(z,\kt) &= -\frac{ e_f e \sqrt{2N_c}}{ (\kt^2 + \epsilon^2)} \left[ ke^{i\theta_k}( z \delta_{h+} \delta_{\bar h -}  \right. \nonumber \\
    & \left. - (1-z)\delta_{h-}\delta_{\bar h +} ) + m\delta_{h+}\delta_{\bar h+}  \right] \\
    \Psi^{\lambda=-1}_{\gamma,h\bar h}(z,\kt) &= -\frac{ e_f e \sqrt{2N_c}}{ (\kt^2 + \epsilon^2)} \left[ ke^{-i\theta_k}( (1-z) \delta_{h+} \delta_{\bar h -}  \right. \nonumber \\
    & \left.  - z \delta_{h-}\delta_{\bar h +} ) +  m\delta_{h-}\delta_{\bar h-} \right]
\end{align}
where $\epsilon^2 = Q^2 z(1-z) + m^2$ and $ke^{i \theta_k} = k_x + ik_y$. 
The wave function in the mixed transverse coordinate, longitudinal momentum fraction space entering in the vector meson production cross section~\eqref{eq:vm_amp} is then obtained by performing a Fourier transform 
\begin{equation}
    \Psi_{\gamma,h\bar h}^\lambda(z,\rt) = \int \frac{\der^2 \kt}{(2\pi)^2} e^{i \kt \cdot \rt} \Psi_{\gamma,h\bar h}^\lambda (z,\kt).
\end{equation}
The mixed space wave function for the longitudinal polarization is
\begin{equation} 
\label{eq:photon_L}
    \Psi_{\gamma,h\bar h}^{\lambda=0}(z,\rt) = -e_f e \sqrt{\nc} \delta_{h,-\bar h} 2 Q z(1-z) \frac{K_0(\epsilon r)}{2\pi}.
\end{equation}
 Similarly, for the transverse photon with $\lambda=\pm 1$ the wave function reads
 \begin{align}
    \Psi^{\lambda=+1}_{\gamma,h\bar h}(z,\rt) &= - e_f e \sqrt{2N_c} \left[ i e^{i\theta_r} \frac{\epsilon K_1(\epsilon r)}{2\pi} ( z \delta_{h+} \delta_{\bar h -}  \right. \nonumber \\
    & \left.  - (1-z)\delta_{h-}\delta_{\bar h +} )  + m \frac{K_0(\epsilon r)}{2\pi} \delta_{h+}\delta_{\bar h+}  \right]  \nonumber \\
    \Psi^{\lambda=-1}_{\gamma,h\bar h}(z,\rt) &= - e_f e \sqrt{2N_c} \left[ ie^{-i\theta_r} \frac{\epsilon K_1(\epsilon r)}{2\pi} ( (1-z) \delta_{h+} \delta_{\bar h -}  \right. \nonumber \\
   &  \left.  - z \delta_{h-}\delta_{\bar h +} ) +  m \frac{K_0(\epsilon r)}{2\pi} \delta_{h-}\delta_{\bar h-} \right].
 \label{eq:photon_T}
\end{align}
We note that the these wave functions agree with those derived in Ref.~\cite{Dosch:1996ss} using the same convention, except for the overall sign in case of transverse polarizations which does not affect any of our results. On the other hand, when compared to the widely used wave functions reported in Ref.~\cite{Kowalski:2006hc}, the relative sign between the mass term and $z$ terms in the $\lambda=+1$ case is different.

We emphasize that the quark light cone helicity structure above does not exactly correspond to the spin structure in the rest frame of the meson (there is no rest frame for the spacelike photon). In particular, when transformed to the meson rest frame, there are both $S$ and $D$ wave contributions in both longitudinally and transversely polarized photons. The transformation between the light front wave function expressed in terms of the quark light front helicities, and the rest frame wave function in terms of the quark spins is discussed in Sec.~\ref{sec:melosh}. We will discuss the decomposition  of light cone wave functions, including the virtual photon  one,  into the $S$ and $D$ wave components in more detail in Appendix~\ref{app:s_d}.

\subsection{Dipole-target scattering}
The dipole-target scattering amplitude $N$ in Eq.~\eqref{eq:vm_amp} is a correlator of Wilson lines, corresponding to the eikonal propagation of the quarks in the target color field. In principle, it satisfies perturbative evolution equations describing the dependence on momentum fraction $\xpom$, the so called JIMWLK equation~\cite{JalilianMarian:1996xn,JalilianMarian:1997jx, JalilianMarian:1997gr,Iancu:2001md, Ferreiro:2001qy, Iancu:2001ad, Iancu:2000hn}, or the BK equation~\cite{Balitsky:1995ub,Kovchegov:1999yj} that is obtained in the large-$\nc$ limit. These perturbative evolution equations, combined with a non-perturbative input obtained by fitting some experimental data, can in principle be used to evaluate the dipole amplitude at any (small) $\xpom$. This has been a successful approach when considering structure functions in DIS or inclusive particle production in hadronic collisions, see e.g. Refs.~\cite{Albacete:2010sy,Albacete:2012xq,Lappi:2012nh,Lappi:2013zma,Ducloue:2015gfa,Ducloue:2016pqr,Mantysaari:2019nnt}. 

In diffractive scattering considered here one explicitly measures the transverse momentum transfer $\Deltat$, which is the Fourier conjugate to the impact parameter. Consequently, the dependence on the transverse geometry needs to be included accurately in the calculation. However, perturbative evolution equations generate long distance Coulomb tails that should be regulated by some non-perturbative physics in order to avoid unphysical growth of the cross section~\cite{GolecBiernat:2003ym}. There have been attempts to include effective confinement scale contributions in the BK and JIMWLK evolutions and use the obtained dipole amplitudes in phenomenological calculations of e.g. vector meson production~\cite{Berger:2011ew,Berger:2012wx,Mantysaari:2018zdd,Bendova:2019psy} (see also~\cite{Schlichting:2014ipa}). As the main focus of this work is in vector meson wave functions, we apply a simpler approach and use the so called IPsat parametrization to describe the dipole-proton scattering amplitude. 

The IPsat parametrization~\cite{Kowalski:2003hm} consist of an eikonalized DGLAP-evolved~\cite{Gribov:1972ri,Gribov:1972rt,Altarelli:1977zs,Dokshitzer:1977sg} gluon distribution, combined with an impact parameter $\bt$ dependent transverse  density profile. The advantage of this parametrization is that it matches perturbative QCD result in the dilute (small dipole size $|\rt|$) limit, and respects unitary in the saturation regime. The dipole amplitude in the IPSat parametrization reads
\begin{equation}
\label{eq:ipsat_p}
    N(\rt,\bt,x) = 1 - \exp \left(-\frac{\pi^2}{2\nc} \rt^2 \as(\mu^2) xg(x,\mu^2) T_p(\bt) \right),
\end{equation}
where the proton transverse density profile is assumed to be Gaussian:
\begin{equation}
    T_p(\bt) = \frac{1}{2\pi B_p} e^{-b^2/(2B_p)}
\end{equation}
with $B=4\gev^{-2}$. The initial condition for the DGLAP evolution is obtained by fitting the HERA structure function data~\cite{Aaron:2009aa,Abramowicz:2015mha,H1:2018flt,Abramowicz:1900rp}, and the fit results in an excellent description of the total reduced cross section and the charm contribution~\cite{Mantysaari:2018nng}. The scale choice is $\mu^2=C/r^2+\mu_0^2$, with the parameters $C$ and $\mu_0$, among with the DGLAP initial condition, are determined in the fit performed in Ref.~\cite{Mantysaari:2018nng} (see also~\cite{Rezaeian:2012ji}).

Following Ref.~\cite{Kowalski:2003hm} (see also \cite{Mantysaari:2018nng}), the dipole-proton scattering amplitude can be generalized to coherent scattering in the dipole-nucleus case as
\begin{equation}
\label{eq:ipsat_a}
    N_A(\rt,\bt,x) = 1 - \exp \left(-\frac{\pi^2}{2\nc} \rt^2 \as(\mu^2) xg(x,\mu^2) A T_A(\bt) \right).
\end{equation}
This estimate is valid in case of large nuclei, assuming that the dipole size $|\rt|$ is not very large, which is the case in heavy vector meson production. Here $T_A(\bt)$ is the Woods Saxon distribution integrated over the longitudinal coordinate, with the normalization $\int \der^2\bt T_A(\bt)=1$. The nuclear radius used here is $R_A = (1.13A^{1/3} - 0.86A^{-1/3})\fm$.

In order to calculate vector meson production, it is still necessary to determine the vector meson wave function. It can not be computed perturbatively, and consequently there are many phenomenological parametrizations used in the literature. 
The main goal of this paper is to obtain the meson wave function in a systematic expansion in quark velocities given by the NRQCD approach. We will also discuss, for comparison, some other wave function parametrizations in  Sec.~\ref{sec:pheno-wf}.

\section{Light cone wave function from NRQCD}
\label{sec:nrqcd}
NRQCD is an effective field theory describing QCD in the limit where quark masses are large, or $v=p/m$ is small, where $p$ is e.g. quark momentum and $m$ is the quark mass. In this approach, it becomes possible to factorize cross sections into universal long distance matrix elements and perturbatively calculated process dependent hard factors. 

\subsection{Vector meson wave function in the rest frame}

The \jpsi decay width in the NRQCD approach is written as an expansion in the quark velocity $v$~\cite{Bodwin:1994jh}. At lowest order in $v$, the decay width is only sensitive to the long distance matrix element $\langle \mathcal{O}_1 \rangle_{\jpsim}$, which itself is determined by  the value of the (renormalized) wave function at the origin. At next order, one finds a  contribution  proportional to the long distance matrix element $\langle {\vec q}^{\, 2} \rangle_{\jpsim}$ which is suppressed by a relative $v
^2$. This matrix element is sensitive to the derivative of the wave function at the origin (see also Ref.~\cite{Braguta:2006wr,Braguta:2007fh} for a discussion of the velocity suppressed contributions to the distribution amplitude).

In this work we follow Ref.~\cite{Bodwin:2007fz}, where these matrix elements are determined. There, a subset of higher order (in $v$) contributions to the decay width including higher powers of $\nabla^2$  are resummed to all orders following Ref.~\cite{Bodwin:2006dn}. As a result, the \jpsi decay width in the leptonic channel can be written as
\begin{multline}
\label{eq:gamma_ee}
    \Gamma(\jpsim \to e^-e^+) =  \frac{8 \pi e_q^2 \alpha_\text{em}^2}{3 M_V^2} \Bigg[1 - f\left(\frac{\langle {\vec q}^{\, 2} \rangle_{\jpsim}}{\mcnr^2} \right)  \\
    - 2 C_\mathrm{F} \frac{\as}{\pi}\Bigg]^2 \langle \mathcal{O}_1 \rangle_{\jpsim}
\end{multline}
with 
\begin{equation}
    f(x)=\frac{x}{3(1+x+\sqrt{1+x})}.
\end{equation}
Here, $e_q=2/3$ is the fractional charge of the charm quark and $M_V$ is the \jpsi\ mass. At this order in $v$, the \jpsi is a pure $S$ wave state, and its wave function can be factorized into a spin part and a scalar part. We will discuss the spin and angular momentum structure in more detail later.

The extraction of the matrix elements that we use~\cite{Bodwin:2007fz} has been done in a calculation that includes both velocity and $\as$ corrections, such as in \eqref{eq:gamma_ee}. Here, on the other hand, we will be using the light cone wave functions in a leading order calculation of cross sections, including only velocity corrections to the wavefunction. In a strict NRQCD power counting sense in $\as$, the $\as$ corrections could be considered more important. Although steps have been taken to take them into account in the dipole picture  exclusive cross section calculations~\cite{Escobedo:2019bxn} (see also recent work in a different formalism~\cite{Flett:2019pux}), fully including them in the cross section is not yet possible at this point since the full photon to heavy quark pair wave function is not known to one loop accuracy. Thus we will leave a computation that includes also the perturbative $\as$ calculations to future work, and continue with our focus on the velocity corrections to the wave function here.

Since our cross section calculation does not include pure $\as$ corrections, taking the wave function to be given by just the operator $\langle \mathcal{O}_1 \rangle_{\jpsim}$ in \eqref{eq:gamma_ee} would lead to an inconsistent treatment of the $\as$ corrections between the decay width and the cross section. 
Even in a more general sense, the $\as$ contributions that appear as corrections to the decay widths or cross sections expressed in terms of nonrelativistic wavefunctions should, in light cone perturbation theory, be thought of as perturbative corrections to the light cone wave function itself~\cite{Frankfurt:1997fj,Escobedo:2019bxn}.
This can be understood in the sense that the degrees of freedom in the nonrelativistic wavefunction are constituent quarks as opposed to bare quarks in the light cone wave function, see the discussion in \cite{Frankfurt:1997fj}. To obtain a consistent picture here, we will absorb the $\as$ correction to the scalar part of the wavefunction $\phi(r)$, which is then transformed to the light cone wave function. We thus relate the value and derivative at the origin of  $\phi(r)$  to the long distance matrix elements as 
\begin{align}
\label{eq:wf_matrix_element_O}
   \Bigg[1 
    - 2 C_\mathrm{F} \frac{\as}{\pi}\Bigg]^2 \langle \mathcal{O}_1 \rangle_{\jpsim} &=2 N_c \lvert \phi(0) \rvert^2 + \mathcal{O}(v^4), \\
\label{eq:wf_matrix_element_q}
   \langle {\vec q}^{\, 2} \rangle_{\jpsim} &= - \frac{\nabla^2 \phi(0)}{\phi(0)} + \mathcal{O}(v^2).
\end{align}

The non-perturbative long distance matrix elements have been determined in  Ref.~\cite{Bodwin:2007fz} by considering simultaneously the $\jpsim \to e^+e^-$ and $\eta_c \to 2\gamma$ decays. As a result of this analysis, the matrix elements for  \jpsi\ read
\begin{align}
\label{eq:matrix_element_O}
   \langle \mathcal{O}_1 \rangle_{\jpsim} &= 0.440^{+0.067}_{-0.055} \, \gev^3, \\
\label{eq:matrix_element_q}
   \langle {\vec q}^{\, 2} \rangle_{\jpsim} &= 0.441^{+0.140}_{-0.140} \gev^2.
\end{align}
The analysis in Ref.~\cite{Bodwin:2007fz} is done by using the charm quark mass $\mcnr=1.4\gev$. In general, the charm quark mass in NRQCD can differ from the charm quark mass used in the IPsat fits discussed in Sec.~\ref{sec:dipolepicture}. In our numerical analysis, we will use the NRQCD value for the charm quark mass in both the meson and photon  wave functions when using the NRQCD results. 
Everywhere else in this work we use the  charm mass $m_c=1.3528 \gev$ obtained in the IPsat fit to the HERA structure function data.

The uncertainties quoted above for the long distance matrix elements are not independent, and the correlation matrix is also provided in Ref.~\cite{Bodwin:2007fz}. To implement these correlated uncertainties, we use a Monte Carlo method and sample parameter values from the Gaussian distribution taking into account the full covariance matrix. The uncertainty is then obtained by calculating the one standard deviation band with respect to the result obtained by using the best fit values.

To construct the meson wave function, we start from the meson rest frame where we can use the NRQCD matrix elements to constrain the wave function as discussed above. In the rest frame, we require that the quark spins are coupled into a triplet state, and the total spin and angular momentum to a $J=1$ vector state. Thus we can in general write the spin-structure of the wave function in the following form:
\begin{multline}
\label{eq:general_wf}
    \psi^\lambda_{s \bar s}(\vec r) = \sum_{L, m_L, m_S} \left\langle L \ m_L \ 1 \ m_S \vert 1 \ \lambda \right\rangle \\
    \times \left\langle \frac{1}{2} \ s  \ \frac{1}{2} \ \bar s \Bigg\vert 1 \ m_S \right\rangle Y_L^{m_L}(\theta, \phi) \psi_L(r).
\end{multline}
 Here $Y_L^{m_L}$ are the spherical harmonics, $\psi_L$ is the radial wave function corresponding to the orbital angular momentum $L$ and $\langle j_1 m_{j_1} j_2 m_{j_2} \vert J m_J \rangle$ are Clebsch-Gordan coefficients. In general, the conservation of spin-parity tells us that for \jpsi the orbital angular momentum can only take values $L = S, D$. Since \jpsi should be dominated by the $S$ wave contribution,  we will from now on consider the case where only the $S$ wave component is non-zero. We note that in principle in the NRQCD approach one finds the $D$ wave contribution to the vector meson wave function to be suppressed by $v
^2$ compared to the $S$ wave, and this is of the same order as the first relativistic correction included in terms of the wave function derivative above. However, the $D$ wave contribution to the decay width is suppressed by an additional $v^2$ and as such the $D$-wave contribution is not constrained by the decay widths at this order. Thus it is most consistent to set it to zero.
In this case the wave function simplifies to: 
\begin{equation}
\label{eq:S-wave}
    \psi^\lambda_{s \bar s}(\vec r) = \left\langle \frac{1}{2} \ s \ \frac{1}{2} \ \bar s \Bigg\vert 1 \ \lambda \right\rangle \phi(r)
\end{equation}
where $\phi(r)$ is the scalar part of the wave function and related to the long distance matrix elements as shown in Eqs.~\eqref{eq:wf_matrix_element_O} and~\eqref{eq:wf_matrix_element_q}. Using the 3-dimensional polarization vectors in Eq.~\eqref{eq:trans_pol} we can also write this as
\begin{equation}
\label{eq:S-wave_var}
    \psi^\lambda_{s \bar s}(\vec r) =  U^\lambda_{s \bar s}\phi(r)
\end{equation}
where 
\begin{equation}
\label{eq:U_spin}
    U^\lambda_{s \bar s} = \frac{1}{\sqrt{2}} \xi^\dag_s \vec \epsilon_\lambda \cdot \vec \sigma \tilde \xi_{\bar s}
\end{equation}
in the case of transverse polarization and
\begin{equation}
\label{eq:U_spin_L}
    U^{\lambda=0}_{s \bar s} = \frac{1}{\sqrt{2}} \xi^\dag_s  \sigma_3 \tilde \xi_{\bar s}
\end{equation}
when the vector meson is longitudinally polarized. Here $\xi_+=(1,0)$ and $\xi_- = (0,1)$ are the two component spinors describing spin-up and spin-down states and $\tilde \xi_{\bar s} = i \sigma_2 \xi_s^*$ is the antiquark spinor. 

The behavior of the quarkonium wavefunction at long distances is determined by non perturbative physics. This long distance physics affects  short distances through the requirement of the normalization of the wave function. The NRQCD approach broadly speaking consists of parametrizing the nonperturbative long distance physics by measurable coefficients that serve as coefficients in the short distance expansion, which is used to calculate a physical process happening at short distance scales. In practice this amounts to expressing the  wave function as a Taylor expansion around the origin:
\begin{equation}
\label{eq:NQRCD_expansion_position_space}
    \phi(\vec r) = A + B \vec r^2.
\end{equation}
The linear term does not appear to ensure that the Laplacian of the wave function is finite at the origin. 
The coefficients can also be written as $ A = \phi(0) $ and $ B = \frac{1}{6} \nabla^2 \phi(0)$, and using equations Eqs.~\eqref{eq:wf_matrix_element_O} and \eqref{eq:wf_matrix_element_q} we get the values 
\begin{align}
\label{eq:A_coefficient}
    A &= \Bigg[1 
    - 2 C_\mathrm{F} \frac{\as}{\pi}\Bigg] \sqrt{\frac{1}{2N_c} \langle \mathcal{O}_1 \rangle_{\jpsim} } = 0.213 \gev^{3/2}, \\ %
\label{eq:B_coefficient}
    B &= - \frac{1}{6} A \langle {\vec q}^{\, 2} \rangle_{\jpsim} =  -0.0157 \gev^{7/2}.
\end{align}
The uncertainties in the long distance matrix elements are correlated as discussed above, and in our numerical calculations this correlated uncertainty is propagated to the coefficients $A$ and $B$.

We then want to write  our wave function ansatz \eqref{eq:NQRCD_expansion_position_space} in light cone coordinates $(\kt, z)$. We do this by first going to momentum space:
\begin{multline}
\label{eq:NQRCD_expansion_momentum_space}
    \psi_{s \bar s}^\lambda (\vec k) = \int \der^3 \vec r e^{-i \vec k \cdot \vec r} \psi_{s \bar s}^\lambda (\vec r) = U^\lambda_{s \bar s} \phi(k)\\
    = U^\lambda_{s \bar s} (2 \pi)^3 \left(A \delta^{3} (\vec k) - B \nabla^2_k \delta^{3} (\vec k) \right)
\end{multline}
where $\vec k = (\kt,k^3)$. We then want to change the longitudinal momentum variable from $k^3$ to the plus momentum fraction carried by the quark: $z$. Unfortunately there is no unique way to do this, due to the different nature of instant form and light cone quantization.  In principle we would want to define $z$ as the ratio of the quark $k^+$ to the meson $P^+= M_V/\sqrt{2}$, working in the rest frame of the meson. However, a quark inside a bound state described as a superposition of different $\vec{k}$ modes is not exactly on shell, its energy being affected by the binding potential. Thus we do not precisely know the $k^0$ required to calculate $k^+$ from $k^3$. The rest frame wave function also includes values of $k^3$ that are very  large, leading to values of $k^+$ that are larger than $M_V/\sqrt{2}$. This is perfectly possible in instant form quantization with the time variable $t$. However, in light cone quantization $k^+$ is a conserved momentum variable, and has to satisfy $0< k^+ < P^+$. The procedure that we adopt here is (similarly to e.g.~\cite{Bodwin:2006dm}) to define the momentum fraction in practice as $z= k^+_q/(k^+_q+ k^+_{\bar{q}})$ where $k_q$ and $k_{\bar q}$ are the quark and antiquark momenta, with $k^+$ calculated assuming $k^0 = \sqrt{\mcnr^2 + \vec{k}^2}$. In other words, we  normalize by the total  plus momentum of the quark-antiquark pair, instead of the meson plus momentum, and assume an on-shell dispersion relation. This choice has the advantage that it leads to $0<z<1$ by construction. 
This leads us to the expression for the longitudinal momentum in the meson rest frame $k^3$ as
\begin{equation}
\label{eq:k_3}
    k^3 =  M \left(z-\frac{1}{2} \right)
\end{equation}
where
\begin{equation}
\label{eq:M}
    M = \sqrt{\frac{\kt^2 +\mcnr^2}{z(1-z)}}
\end{equation}
is the invariant mass of the quark-antiquark pair. 
We emphasize that since this choice is not unique, we might expect corrections or ambiguities proportional to powers of the difference between the meson mass and the quark-antiquark pair invariant mass 
$M_V^2 - M^2$ to appear. Such corrections are, however, higher order corrections in the nonrelativistic limit and also numerically very small for \jpsi for the values of $\mcnr$ and $\langle {\vec q}^{\, 2}\rangle$ used here. We could also hope that since the invariant mass is a rotationally invariant quantity, these ambiguities would not lead to serious violations of rotational invariance (which expresses itself here as the equality of physical properties such as decay widths of transverse and longitudinal polarization states). We will see an example of such a correction explicitly in Appendix~\ref{app:scalarlcwf}.

To change the variables in our wave function, one needs to be careful with the delta functions and their derivatives. We therefore make the change by requiring that the overlap \begin{equation}
\label{eq:change_of_variables}
    \int \frac{\der^3 \vec k}{(2 \pi)^3} \psi_{s \bar s}^\lambda (\vec k) \varphi(\vec k) = \int \frac{\der^2 \kt}{(2\pi)^2} \frac{\der z}{4\pi}  \psi_{s \bar s}^\lambda (\kt, z) \varphi(\kt, z),
\end{equation}
where $\varphi$ is an arbitrary wave function, does not change under the change of variables. This requirement tells us that the scalar part $\phi(\vec k)$ changes to
\begin{multline}
\label{eq:phi_kt_z}
    \phi(\kt,z) =  (2 \pi)^3 \sqrt{2 \frac{\partial z}{\partial k^3}} \left(A\delta \left(z -\frac{1}{2} \right)\delta^2(\kt) \right. \\
    \left. - B\left(\partial_z \left[ \frac{\partial z}{\partial k^3} \partial_z \left[\frac{\partial z}{\partial k^3} \delta\left(z - \frac{1}{2}\right) \right] \right]\delta^2(\kt) \right. \right. \\
    \left.\left.+ \delta\left(z - \frac{1}{2}\right) \nabla_{\kt}^2 \delta^2(\kt) \right) \right)
\end{multline}
where
\begin{equation}
\label{eq:dzdk3}
    \frac{\partial z}{\partial k^3} = \frac{4 z (1-z)}{M}.
\end{equation}

Equation \eqref{eq:phi_kt_z} is the scalar part of the NRQCD based vector meson wave function in the meson rest frame, expressed in momentum space. We note that this wave function is not normalizable due to the presence of the delta functions. However, as the NRQCD approach can only be used to constrain the coordinate space wave function and its derivative at the origin, we are forced to use the expansion of Eq.~\eqref{eq:NQRCD_expansion_position_space} which can not result in a normalizable wave function.
However, for the purposes of this work this is not a problem, as the vector meson production is sensitive to the vector meson wave function overlap with the virtual photon wave function, and the photon wave function is heavily suppressed at long distances where the expansion~\eqref{eq:NQRCD_expansion_position_space} is not reliable.

\subsection{Wave function on the light front}
\label{sec:melosh}
The NRQCD  wave function obtained in the previous section is written in the vector meson rest frame in terms of the quark and antiquark spin states $s,\bar s$. In order to calculate overlaps with the virtual photon wave function \eqref{eq:photon_L} and \eqref{eq:photon_T}, we need to express it in terms of the light cone helicities $h,\bar h$. The transformation between these two bases, usually expressed in terms of the 2-spinors, is known as the  ``Melosh rotation''~\cite{Melosh:1974cu,Hufner:2000jb}.

The Dirac spinors that are used to factorize the nonrelativistic wavefunction into a a spin- and scalar part, are eigenstates of the spin-$z$ operator in the zero transverse momentum limit. In terms of the two component spin vectors $\xi$ defined above in \eqs\eqref{eq:U_spin} and~\eqref{eq:U_spin_L} they read
\begin{align}
\label{eq:spin-u}
    u_s(p) &= \frac{1}{\sqrt{N}}\begin{pmatrix} \xi_s \\ \frac{\vec \sigma \cdot \vec p}{E_p+m} \xi_s \end{pmatrix} \\
    \label{eq:spin-v}
    v_s(p) &= \frac{1}{\sqrt{N}}\begin{pmatrix} \frac{\vec \sigma \cdot \vec p}{E_p+m} \tilde \xi_{s} \\  \tilde \xi_{s} \end{pmatrix}.
\end{align}
The normalization factor $N$ is determined from the condition $\bar u_s u_{\bar s} = -\bar v_s v_{\bar s} = 2m \delta_{s,\bar s}$. 

 Both the Dirac spinors in terms of the spin-$z$ component $u_s$ and the helicity spinors $u_h$ (see \eqs\eqref{eq:uh}, \eqref{eq:vh}) are solutions to the Dirac equation, and as such can be obtained as linear combinations of each other. This mapping is the Melosh rotation $R^{sh}$. It can be computed from the spinor inner products (see also Ref.~\cite{Krassnigg:2001ka}) as
\begin{equation}
    R^{sh}(\kt,z) = \frac{1}{2m} \bar u_s(\kt,z) u_h(\kt,z)
\end{equation}
where $k^+ = zq^+$ and $q^+$ is the meson plus momentum, and  $s$ and $h$ refer to the spin and light front helicity, respectively. 

The helicity spinors $u_h$ and $v_h$ can also be written in a similar form as the spinors in the spin basis, Eqs.~\eqref{eq:spin-u} and \eqref{eq:spin-v}, by introducing the two component helicity spinors $\chi_h$. To do this we  write the helicity spinors \eqref{eq:uh}, \eqref{eq:vh} in the form
\begin{align}
\label{eq:spin-uh}
    u_h(p) &= \frac{1}{\sqrt{N}}\begin{pmatrix} \chi_h \\ \frac{\vec \sigma \cdot \vec p}{E_p+m} \chi_h \end{pmatrix} \\
    \label{eq:spin-vh}
    v_h(p) &= \frac{1}{\sqrt{N}}\begin{pmatrix} \frac{\vec \sigma \cdot \vec p}{E_p+m} \tilde \chi_{h} \\  \tilde \chi_{h} \end{pmatrix} 
\end{align}
where $N$ is again determined by the normalization requirement and $\tilde \chi_h = i \sigma_2 \chi_h^*$. Using this form one can check that the Melosh rotation also connects the two component spin and helicity spinors as
\begin{equation}
\label{eq:melsoh_2comp}
    R^{sh}(\kt,z)= \xi_s^\dagger \chi_h
\end{equation}

The  coefficients $R^{sh}$ can also be expressed as a 2$\times$2 matrix rotating the 2-spinors 
\begin{equation}
\label{eq:melosh_r}
    R(\kt,z) = \frac{m_{c,NR} + zM - i(\vec \sigma \times \vec n) \cdot (\kt,k^3) }{\sqrt{(m_{c,NR}+ zM)^2 + \kt^2}}.
\end{equation}
Here $M$ is the invariant mass of the $q\bar q$ system from Eq.~\eqref{eq:M} and  $n=(0,0,1)$ is the unit vector in the longitudinal direction. In terms of this matrix the 2-spinors $\xi_s$ and $\chi_h$ are related by
\begin{equation}
\chi_\pm = R(\kt,z) \xi_\pm
\end{equation}
Using Eq.~\eqref{eq:melsoh_2comp} we can now express the NRQCD wave function in the light front helicity basis. We write
\begin{equation}
    \Psi_{h\bar h}^\lambda (\kt, z) = U^\lambda_{h,\bar h} \phi(\kt, z),
\end{equation}
where the scalar part is given in Eq.~\eqref{eq:phi_kt_z}. The helicity structure $U_{h\bar h}^\lambda$ is obtained by applying the transform \eqref{eq:melsoh_2comp} in Eq.~\eqref{eq:U_spin} and \eqref{eq:U_spin_L}, i.e.,
\begin{equation}
\label{eq:U_helicity}
    U_{h \bar h}^\lambda = \sum_{s \bar s} R^{*sh}(\kt,z) R^{*\bar s \bar h}(-\kt, 1-z) U_{s\bar s}^\lambda .
\end{equation}

After the Melosh rotation, we compute the Fourier transform to obtain the light front wave function in the mixed transverse coordinate -- longitudinal momentum fraction space as
\begin{multline}
    \Psi_{h\bar h}^\lambda(\rt, z) = \int \frac{\der^2 \kt}{(2 \pi)^2}e^{i \kt \cdot \rt} \Psi_{ h\bar h}^\lambda(\kt, z)  \\
    =  \int \frac{\der^2 \kt}{(2 \pi)^2}e^{i \kt \cdot \rt} U^\lambda_{h,\bar h}(\kt, z) \phi(\kt, z).
\end{multline}
The different helicity components of the final light front wave function resulting from this procedure are
\begin{widetext}
\begin{align}
\nonumber
        \Psi^{\lambda = 0}_{+-}(\rt,z) & = \Psi^{\lambda = 0}_{-+}(\rt,z) = \frac{\pi \sqrt{2}}{\sqrt{m_{c,NR}}} \Bigg[A \delta(z-1/2)
            +\frac{B}{m_{c,NR}^2} \Bigg( \left(\frac{5}{2}+\rt^2 m_{c,NR}^2\right) \delta(z-1/2) 
            - \frac{1}{4}\partial_z^2 \delta(z-1/2)\Bigg) \Bigg] 
            \\ \nonumber
            \Psi^{\lambda = 1}_{++}(\rt,z) &= \Psi^{\lambda = -1}_{--}(\rt,z) = \frac{2 \pi}{\sqrt{m_{c,NR}}} \Bigg[A \delta(z-1/2)
            +\frac{B}{m_{c,NR}^2} \Bigg( \left(\frac{7}{2}+\rt^2 m_{c,NR}^2\right) \delta(z-1/2) 
            - \frac{1}{4}\partial_z^2 \delta(z-1/2) \Bigg) \Bigg]
            \\ \nonumber
            \Psi^{\lambda = 1}_{+-}(\rt,z) &= -\Psi^{\lambda = 1}_{-+}(\rt,z)
            =\left(\Psi^{\lambda = -1}_{-+}(\rt,z) \right)^*= \left(-\Psi^{\lambda = -1}_{+-}(\rt,z)\right)^*            
            = -\frac{2 \pi i}{m_{c,NR}^{3/2}} B \delta(z-1/2) (r_1 + i r_2)  
            \\
            \Psi^{\lambda = 1}_{--}(\rt,z) &= \Psi^{\lambda = -1}_{++}(\rt,z)=
            \Psi^{\lambda = 0}_{++}(\rt,z) = \Psi^{\lambda = 0}_{--}(\rt,z) =
            0 
        \label{eq:jpsi_nrqcd}
\end{align}
\end{widetext}

The first relativistic correction to the wave function, proportional to $B$ or the wave function derivative, mixes the helicity and spin states. In particular, in the case of transverse polarization the $h,\bar h = \pm \mp$ terms are non-vanishing when the relativistic correction is included. These terms also bring a non-zero contribution to photon-vector meson overlaps. In general, we expect that if higher order corrections in $v$ were included in the wave function parametrization, we would also find other components to be non-vanishing.

The Melosh rotation is crucial here, as it generates helicity structures that are not visible in the spin basis. This is in contrast to some early attempts to transform the wave functions obtained by solving the potential models to the light front as done e.g. in Ref.~\cite{Kowalski:2003hm}. The role of the Melosh rotation in the context of vector meson light front wave functions and exclusive scattering was first emphasized in Ref.~\cite{Hufner:2000jb}. More recently it was applied to  \jpsi production in the dipole picture in Ref.~\cite{Krelina:2019egg}, and in \cite{Cepila:2019skb} different quark-antiquark potentials were studied in this context. In the case of excited states such as $\psi(2S)$ the role of the Melosh rotation is expected to be even more significant~\cite{Krelina:2018hmt}.

Let us in passing briefly compare our approach to the one in the recent work of Krelina et al in Ref.~\cite{Krelina:2019egg}. In our approach, we take the NRQCD wave function which only includes the $S$ wave contribution ($D$ wave part is suppressed by $v^2$). The quark spin dependence is now trivial, as the total angular momentum must be provided by the quark spins which gives us the structure of Eq.~\eqref{eq:S-wave_var}. In Ref.~\cite{Krelina:2019egg}, the authors assume, unlike we do here, that the spin structure of the   vector meson wave function in the rest frame has the same form as the light cone helicity structure of the photon light cone wave function, Eq.~\eqref{eq:photon_L} and \eqref{eq:photon_T}. This structure is then supplemented by a wave function obtained from the potential model, and a Melosh rotation to the light front is applied at the end. Such a procedure leads to a large $D$-wave contribution in the wavefunction, which we do not have. We discuss the structure of the wavefunctions in terms of $S$- and $D$-waves in more detail in Appendix~\ref{app:s_d}.

To determine the role of the relativistic corrections in the vector meson wave function, we will also study for comparison the fully non-relativistic wave function where our starting point for the scalar part is 
\begin{equation}
    \label{eq:delta_wf_scalar}
    \phi(\vec r) = A'.
\end{equation}
Following the previous procedure, the final result for the light cone wave function can be read from Eq.~\eqref{eq:jpsi_nrqcd} with the substitutions $A=A', B=0$. One notices that this can now be written as
\begin{equation}
\label{eq:deltawf}
    \Psi_{\jpsim,h\bar h}^\lambda(\rt,z) = \frac{\pi \sqrt{2}}{\sqrt{\mcnr}} U^\lambda_{h\bar h}  A'\delta\left(z-\frac{1}{2}\right).
\end{equation}
In this extreme non-relativistic limit $(\kt = 0, z= 1/2)$ the Melosh rotation simply corresponds to an identity matrix so that the spin and helicity bases are interchangeable here. 
The normalization $A'$ is obtained from the van Royen-Weisskopf equation for the leptonic width~\cite{VanRoyen:1967nq}, which is also obtained from Eq.~\eqref{eq:gamma_ee} by neglecting the relativistic correction proportional to $\langle \mathbf{q}^2 \rangle_{\jpsim}/\mcnr^2$, and the higher order QCD correction $\sim \as$ (note that parametrically $\as \sim v$):
\begin{equation}
\label{eq:van_Royen-Weisskopf}
    \Gamma(\jpsim \rightarrow e^- e^+) = \frac{16 \pi e_f^2 \alpha_\textrm{em}}{M_{\jpsim}^2} \lvert \phi(0) \rvert^2
\end{equation}
By using the experimental value for leptonic width~\cite{Tanabashi:2018oca}, we can calculate the coefficient $A'$ to be
\begin{equation} \label{eq:Aprimevalue}
    A'= \phi(0) = 0.211\gev^{3/2}.
\end{equation}

\subsection{Overlap with photon}
Using the obtained \jpsi wave function on the light front, Eq.~\eqref{eq:jpsi_nrqcd}, we can directly compute overlaps with the virtual photon, Eqs.~\eqref{eq:photon_L} and \eqref{eq:photon_T}. In these overlaps, we also include the phase factor $\exp\left(i\left(z-\frac{1}{2}\right)\rt \cdot\Deltat \right)$ present in the vector meson production amplitude in Eq.~\eqref{eq:vm_amp}. We also assume that the dipole amplitude does not depend on the orientation $\theta_r$ of $\rt$ as is the case in the IPsat parametrization, and integrate over $\theta_r$. 
The overlaps summed over the quark helicities 
read
\begin{widetext}
    \begin{multline}
        r\sum_{h \bar h} \int_0^{2 \pi} \der \theta_r\int_0^1 \frac{\der z}{4\pi} (\Psi^L_{\jpsim})^*\Psi^L_{\gamma} e^{i (z-1/2) \rt \cdot \Deltat}%
        = -\frac{r e e_f Q}{2} \sqrt{\frac{N_c}{2\mcnr}}\Bigg[A K_0(r\bar \epsilon) 
\\
        +\frac{B}{\mcnr^2} \bigg(\frac{9}{2} K_0(r\bar \epsilon)
        +\mcnr^2r^2 K_0(r \bar \epsilon) 
        - \frac{Q^2r}{4 \bar \epsilon} K_1(r \bar \epsilon)+\frac{1}{8} \Delta^2 r^2 K_0(r \bar \epsilon) \bigg) \Bigg]
    \end{multline}
    and
    \begin{multline}
        r \sum_{h \bar h}\int_0^{2 \pi} \der \theta_r \int_0^1 \frac{\der z}{4\pi}  (\Psi^T_{\jpsim})^*\Psi^T_{\gamma} e^{i (z-1/2) \rt \cdot \Deltat} %
        =-r e e_f \sqrt{\frac{N_c m_{c,NR}}{2}} \Bigg[A K_0(r\bar \epsilon)
\\
        +\frac{B}{\mcnr^2} \bigg(\frac{7}{2} K_0(r\bar \epsilon)
        +\mcnr^2r^2 K_0(r \bar \epsilon) 
         - \frac{r}{2 \bar \epsilon} (Q^2+2 \mcnr^2) K_1(r \bar \epsilon)
         +\frac{1}{8} \Delta^2 r^2 K_0(r \bar \epsilon)  \bigg) \Bigg],
    \end{multline}
    \end{widetext}
    where $\bar \epsilon^2 = Q^2/4+\mcnr^2$ and $\Delta=|\Deltat|$ and $r=|\rt|$. In the case of transverse polarization, the result is identical in cases with $\lambda=+1$ and $\lambda=-1$.
    We will study these overlaps numerically in Sec.~\ref{sec:numerics_wf}.
    We note that thanks to the delta function structure in $z$ in our wave function \eqref{eq:jpsi_nrqcd}, many phenomenological applications become numerically more straightforward as the $z$ integral can be performed analytically.

\section{Phenomenological wave functions}
\label{sec:pheno-wf}

To provide a quantitative point of comparison for the effect of the relativistic corrections, we want to compare the light cone wave functions obtained in Sec.~\ref{sec:nrqcd} to other parametrizations used in the literature. For this purpose, let us now discuss two specific alternative approaches used for phenomenological applications in the literature.

\subsection{Boosted Gaussian}
\label{sec:boosted_gaussian}
A commonly used phenomenological approach to construct the vector meson wave function is to assume that it has the same polarization and helicity structure as the virtual photon. This can be done by replacing the scalar part of the photon wave functions \eqref{eq:photon_L} and \eqref{eq:photon_T} 
by an unknown function as~\cite{Kowalski:2006hc}
\begin{equation}
    e_f e z(1-z) \frac{K_0(\epsilon r)}{2\pi} \to \phi_{T,L}(r,z),
\end{equation}
with the explicit factor $Q$ in the longitudinal wave function  replaced by the meson mass as $2Q \to M_V$.
The scalar function $\phi(r,z)$ is then parametrized, and the parameters can be determined by requiring that the resulting wave function is normalized to unity and reproduces the experimental leptonic decay width. As we will discuss in more detail in Appendix~\ref{app:scalarlcwf}, this procedure does not correspond to the most general possible helicity structure. Nevertheless, our result at this order in the nonrelativistic expansion can in fact also be written in terms of the ``scalar part of light cone wave functions.'' However,  at higher orders in $v$ different a different structure could appear.

In the Boosted Gaussian parametrization, the $q\bar q$ invariant mass distribution is assumed to be Gaussian, with the width of the distribution $\mathcal{R}$ and the normalization factors $N_{T,L}$ being free parameters. In mixed space, the parametrization reads
\begin{multline}
    \phi_{T,L}(r,z) = \mathcal{N}_{T,L} z(1-z) \exp \left( -\frac{m_c^2 \mathcal{R}^2}{8z(1-z)} \right. \\ 
   \left. - \frac{2z(1-z)r^2}{\mathcal{R}^2} + \frac{m_c^2 \mathcal{R}^2}{2} \right).
\end{multline}
In this work we use the parameters constrained in Ref.~\cite{Mantysaari:2018nng} by using the same charm quark mass $m_c=1.3528 \gev$ as is used when fitting the IPsat dipole amplitude to the HERA data. The parameters are determined by requiring that the longitudinal polarization can be used to reprodcue the experimental decay width. The obtained parameters are $\mathcal{R}=1.507\gev^{-1}$, $N_T=0.589$ and $N_L=0.586$ with $M_V=3.097\gev$.

The specific functional form and helicity structure of the Boosted Gaussian parametrization imply that in the vector meson rest frame there are both $S$ and $D$ wave contributions. This is demonstrated explicitly in Appendix~\ref{app:s_d} by performing a Melosh rotation from the light front back to the \jpsi rest frame. This feature is hard to describe in potential model calculations, and our NRQCD based wave function in particular has only the $S$ wave component in the rest frame. The $D$-wave contribution in the Boosted Gaussian wavefunction is, however, quite small.

\subsection{Basis Light-Front Quantization (BLFQ)}
\label{sec:blfq}
The second wave function we study here for comparisons is based on explicit calculations on the light front. In this approach, one constructs a light front Hamiltonian $H_\text{eff}$, which consists of a one gluon exchange interaction, and a non-perturbative confining potential inspired by light-front holography. The formalism is developed in Refs.~\cite{Vary:2009gt,Honkanen:2010rc,Zhao:2014xaa,Wiecki:2014ola,Adhikari:2016idg,deTeramond:2008ht,Brodsky:2014yha}. %

The quarkonium states are  obtained by solving the eigenvalue problem
\begin{equation}
    H_\text{eff} \left|\psi_{m_J}^{J^{PC}}\right\rangle = M_V^2 \left|\psi_{m_J}^{J^{PC}}\right\rangle.
\end{equation}
As a solution, one obtains the invariant mass $M_V^2$ spectrum and the light front wave functions in momentum space
\begin{equation}
    \psi_{m_J}^{J^{PC}}(\kt, z, h, \bar h) = \left\langle \kt,z,h,\bar h\left|\psi_{m_J}^{J^{PC}}\right\rangle\right.
\end{equation}
Here $J,P,C$ and $m_J$ are the total angular momentum, parity, $C$-parity and the magnetic quantum number of the state, respectively. 
The free paramters, value of the coupling constant, strength of the confining potential,  quark mass and the effective gluon mass, can be constrained by the   charmonium and bottonium mass spectra~\cite{Li:2015zda,Li:2017mlw}. In this work, we use the most up-to-date parametrizations from Ref.~\cite{Li:2017mlw}. 

The obtained BLFQ wave functions have been applied in studies of the \jpsi\ production in the dipole picture at HERA~\cite{Chen:2016dlk}, and in the context of exclusive \jpsi\ production in ultra peripheral heavy ion collisions at the LHC in Ref.~\cite{Chen:2018vdw}. Following the prescription used in Refs.~\cite{Chen:2016dlk,Chen:2018vdw}, we consider the fitted quark mass $m_c^\textbf{BLFQ}=1.603\gev$ in the ``BLFQ wave function'' to be an effective mass of the quarks in the confining potential, including some non-perturbative contributions. Consequently, when calculating the overlaps we use, as in~\cite{Chen:2016dlk,Chen:2018vdw}, $m_c=1.3528\gev$ for the charm mass in the photon wave function, as constrained by the charm structure function data in the IPsat fit~\cite{Mantysaari:2018nng}.

\section{Vector meson production}
\label{sec:numerics}

\subsection{Photon overlap}
\label{sec:numerics_wf}
\begin{figure*}[tb]
\includegraphics[width=\textwidth]{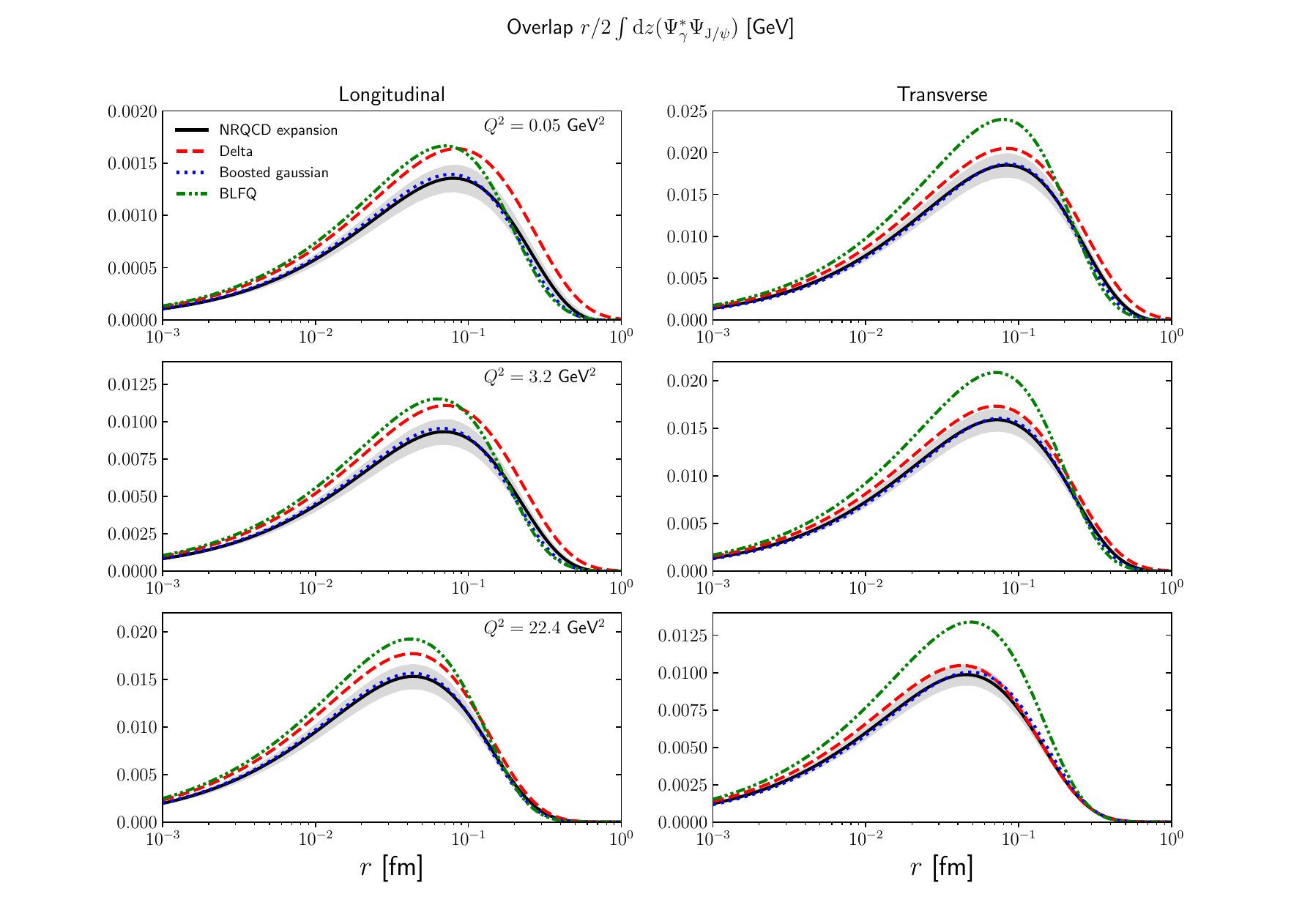}
\caption{Forward ($\Delta=0$) virtual photon-\jpsi wave function overlaps computed using the different vector meson wave functions as a function of the dipole size $r$ at different photon virtualities.
}
\label{fig:overlaps}
\end{figure*}

\begin{figure*}[tb]
\includegraphics[width=\textwidth]{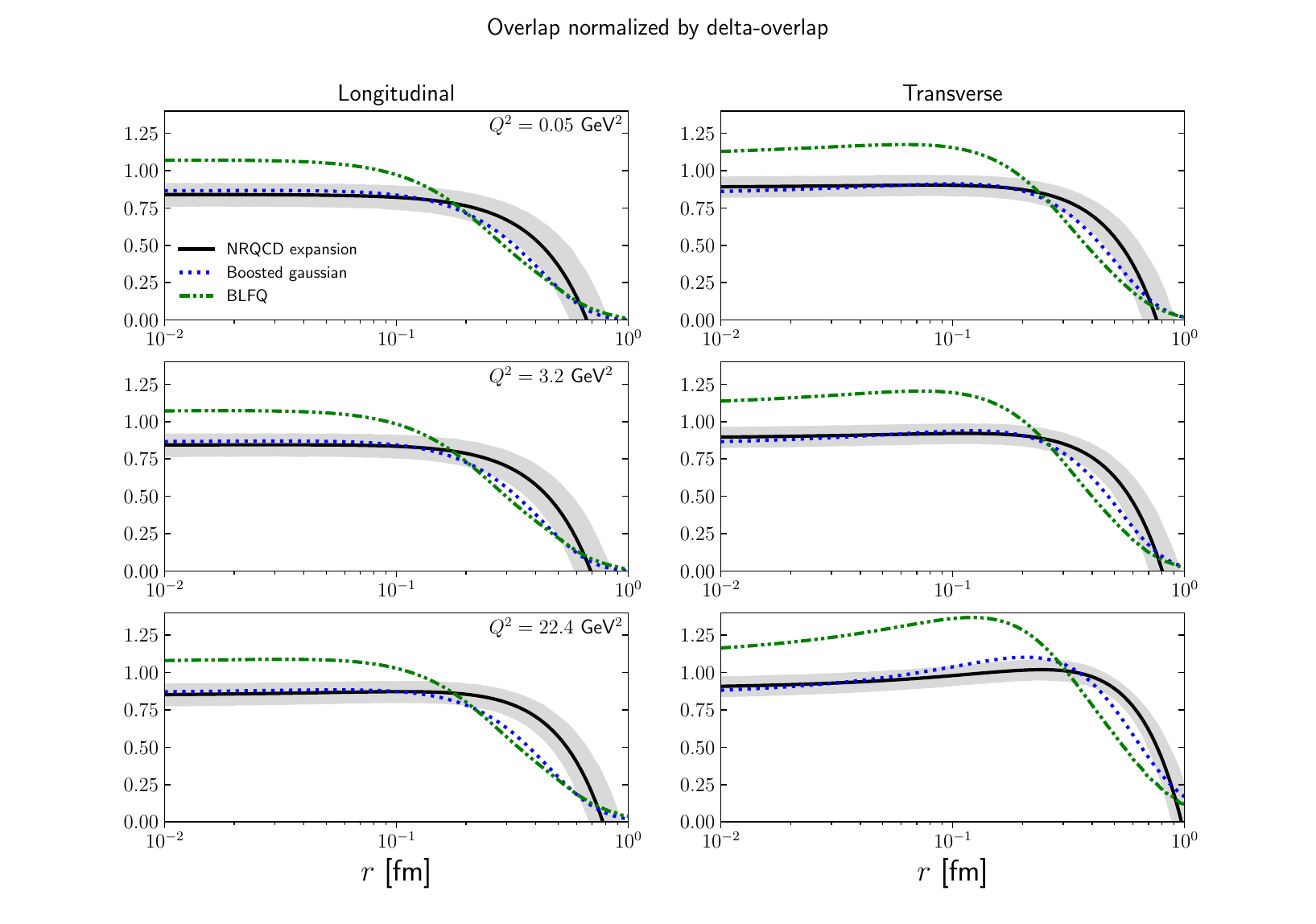}
\caption{Ratios of the forward ($\Delta=0$) virtual photon-\jpsi\ wave function overlaps computed using the different vector meson wave functions to the fully nonrelativistic \emph{Delta} parametrization as a function of the dipole size $r$ at different photon virtualities.
}
\label{fig:overlap_ratio}
\end{figure*}

The exclusive vector meson production cross section depends on the overlap between the $c\bar c$ component of the virtual photon wave function with the vector meson wave function, see Eq.~\eqref{eq:vm_amp}. In Fig.~\ref{fig:overlaps}  these overlaps for $\Delta=0$ are shown as a function of the transverse size $r=|\rt|$ of the intermediate dipole, using four vector meson wave functions:
\begin{enumerate}
    \item \emph{NRQCD expansion}, which is constructed by parametrizing the wave function and its derivative at the origin based on NRQCD matrix elements including corrections $\sim v^2$, and performing the Melosh rotation to the light front. This is our result from Sec.~\ref{sec:nrqcd}. 
    \item \emph{Delta}, which is the fully non-relativistic limit (Eq.~\eqref{eq:deltawf}) of the above wave function, without any information about the wave function derivative.
    \item \emph{Boosted Gaussian}, the phenomenological parametrization discussed in Sec.~\ref{sec:boosted_gaussian}
    \item \emph{BLFQ} wave function based on Basis Light-Front Quantization, discussed in Sec.~\ref{sec:blfq}.
\end{enumerate}
In Fig.~\ref{fig:overlap_ratio} we show the same overlaps plotted as ratios to the fully nonrelativistic limit, i.e. the \emph{Delta} parametrization.
For the NRQCD expansion based wave function, we also show the model uncertainty related to the NRQCD matrix elements that control the value of the wave function and its derivative at the origin. The uncertainty band is in this case computed as discussed in Sec.~\ref{sec:nrqcd}.

The effect of the first relativistic correction can be determined by comparing the \emph{Delta} and \emph{NRQCD expansion} wave functions. At large dipoles the negative velocity suppressed $\sim \rt^2$ contribution suppresses the vector meson wave function\footnote{The wave function would change sign at $r_0=0.73\fm$. As there should be no node in the \jpsi\ wave function, we set the wave function to zero at $r>r_0$. We have checked that this cutoff has a negligible effect on our numerical results.} compared to the fully non-relativistic form. This is especially visible at small $Q
^2$. At larger photon virtualities, the exponential suppression in the photon wave function becomes dominant before the relativistic $ -\rt^2$ correction becomes numerically important. Thus, while the effect of the relativistic correction is dramatic in the ratio in Fig.~\ref{fig:overlap_ratio}, at large $Q^2$ it is insignificant for the actual overlap, as is seen in Fig.~\ref{fig:overlaps}. 

For small dipoles the wavefunctions are most strictly constrained by the quarkonium decay widths. The NRQCD parametrization does not, however, reduce exactly to the fully nonrelativistic \emph{Delta} parametrization in the small $r$ limit. This can be traced back to the fact that the gradient correction also affects the decay width, as seen in \eq\eqref{eq:gamma_ee} (and from the fact that the constants $A$ in \eqref{eq:A_coefficient}  and $A'$ in \eqref{eq:Aprimevalue} are different). A part of the 3-dimensional gradient correction becomes a correction to the functional form in $z$ even at $\rt=0$. This leads to the overlaps at small $r$ being slightly different, even though the same decay width data is used to obtain the parameters of the rest frame wavefunctions.

Both the Boosted Gaussian and BLFQ wave functions are even more suppressed at large dipole sizes than the NRQCD parametrization. This is most clearly seen on the ratio plot, Fig.~\ref{fig:overlap_ratio}. This is a straightforward consequence of the fact that in these parametrizations the wavefunction normalization imposes an additional suppression at large $r$. For the Boosted Gaussian parametrization this additional suppression happens at such a large $r$ that the overlap is already very small, and thus has a negligible effect on the overall overlap in Fig.~\ref{fig:overlaps}.
The Boosted Gaussian parametrization is very close to our NRQCD also for small dipoles. The  BLFQ parametrization yields a a somewhat larger wave function overlap at small $r$ than our NRQCD one, or the Boosted Gaussian.\footnote{The parameters in the BLFQ wave function are constrained by the charmonium mass spectrum, and not the decay widths that probe the wave function at $r=0$. Consequently the BLFQ wave function is not required to result in exactly the same decay width as the other wave functions, which explains the difference at small $r$.}

The suppression with respect to the nonrelativistic limit is larger for the longitudinal polarization state than for the transverse one.  This can be understood as follows. The longitudinal virtual photon wave function depends on the quark momentum fraction as $\sim z(1-z)$ (see Eq.~\eqref{eq:photon_L}), and as  such is peaked at $z=1/2$. The $z$-structure of the fully non-relativistic wave function is $\delta(z-1/2)$, and when the first relativistic corrections are included, the $z=1/2$ region still dominates the overlap.
On the other hand, the transverse photon wave function is not peaked at $z=1/2$, see Eq.~\eqref{eq:photon_T}. Thus, the suppression from the $\partial_z^2\delta(z-1/2)$ term in the relativistic correction is smaller for the transverse polarization.

\subsection{\jpsi\ production}
\begin{figure}
    \includegraphics[width=0.5\textwidth]{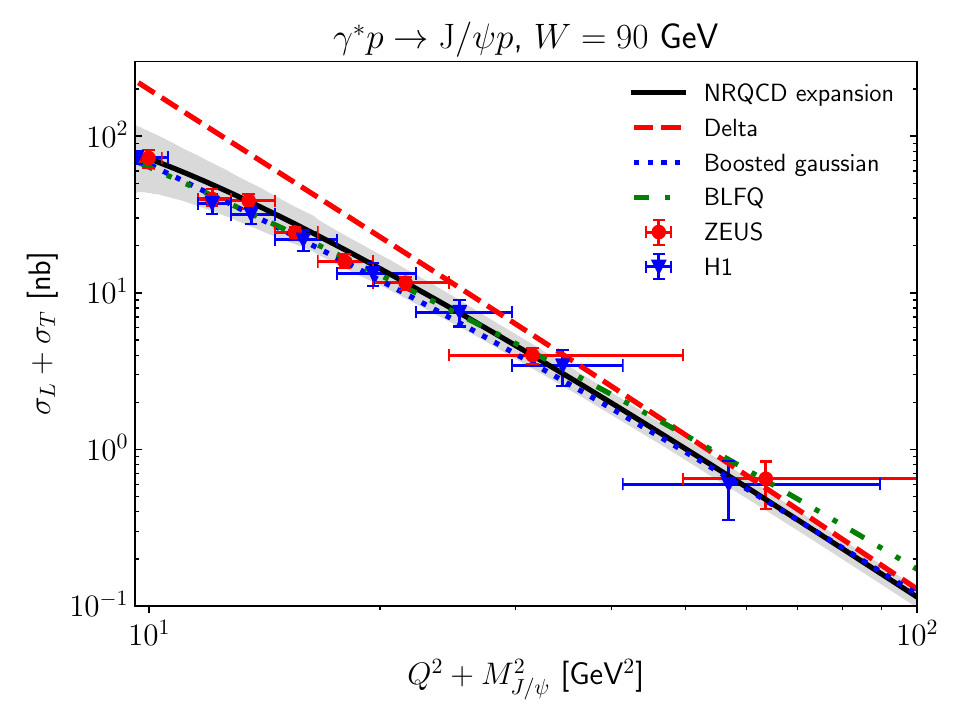}
    \caption{Total \jpsi\ production cross section as a function of virtuality computed using different vector meson wave functions compared with H1~\cite{Aktas:2005xu} and ZEUS~\cite{Chekanov:2004mw} data. 
    }
    \label{fig:totxs_jpsi}
\end{figure}

The total exclusive DIS \jpsi production cross section for a proton target at $W=90$GeV is shown in Fig.~\ref{fig:totxs_jpsi}, compared with the H1~\cite{Aktas:2005xu} and ZEUS~\cite{Chekanov:2004mw} data. 
The overall normalization of the cross section has a relatively large theoretical  uncertainty. 
We note that the two corrections discussed in Sec.~\ref{sec:dipolepicture}, the real part and especially the skewedness correction are numerically significant, up to $\sim 50\%$ (see e.g. Ref.~\cite{Mantysaari:2017dwh}). As discussed in Sec.~\ref{sec:dipolepicture}, especially the skewedness correction is not very robust and its applicability in the dipole picture used here is not clear.
In addition to the possibly problematic skewedness corrections, the fact that our NRQCD based wave functions are not normalized affects the absolute normalization of the vector meson production cross sections.
Thus our focus here is rather on the relative effects of different meson wave functions and the dependence on $Q^2$.

The vector meson cross section is dominated by dipole sizes of the order of $1/(Q^2+M_V^2)$ as can be seen\footnote{Note, however, that as the dipole amplitude scales as $N\sim r^2$ at small $r$, the dominant dipole size scale for the cross section is larger than the maximum of the overlap peaks in Fig.~\ref{fig:overlaps}.} from  Fig.~\ref{fig:overlaps}. Consequently, it is more instructive to look at the dependence of the  \jpsi cross section  on $Q^2$ than the overall normalization.
From Fig.~\ref{fig:totxs_jpsi} one sees that the fully nonrelativistic wave function results in a too steep $Q^2$ dependence compared to the  HERA data.
The first relativistic correction slows down the $Q^2$ evolution close to the photoproduction region and leads to a better agreement with the experimental data. This is a consequence of the basic behavior of the relativistic correction as a $\sim -\rt^2$ modification that suppresses the vector meson wave function strongly at large dipoles. Thus the reduction from the relativistic correction is larger for smaller $Q^2$.
At large $Q^2$ the exponential suppression from the photon wave function starts to dominate at smaller dipole sizes, and the relativistic $-\rt^2$ correction becomes negligible. However, the relativistic contribution to the momentum fraction $z$ structure is present at all $Q^2$, and suppresses the longitudinal cross section more than the transverse one. 

A similar trend in the $Q^2$ dependence is also visible with both the Boosted Gaussian and BLFQ wave functions.
For the Boosted Gaussian case, the agreement with HERA data has been established numerous times in the previous literature, e.g. in Ref.~\cite{Kowalski:2006hc}. The $Q^2$ dependence of the cross section is slightly weaker when the BLFQ wave function is used, but the difference is comparable to the experimental uncertainties.
We note that in Ref.~\cite{Chen:2016dlk} the BLFQ wave function is found to result in a cross section underestimating the HERA data in the photoproduction region. In this work, compared to the setup used in Ref.~\cite{Chen:2016dlk}, we use an updated BLFQ parametrization from Ref.~\cite{Li:2017mlw} which was shown in Ref.~\cite{Chen:2018vdw} to result in a good description of the \jpsi production in ultra peripheral proton-proton collisions at the LHC, which in practice probe vector meson photoproduction~\cite{Bertulani:2005ru,Klein:2019qfb}.

\begin{figure}
    \includegraphics[width=0.5\textwidth]{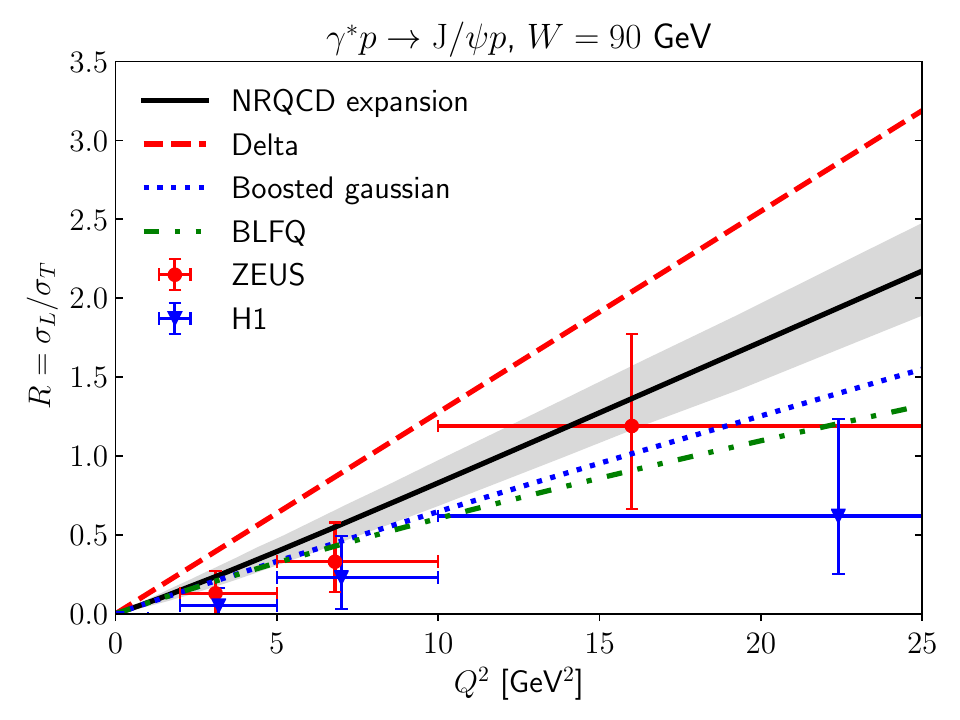}
    \caption{Longitudinal \jpsi\ production cross section divided by the transverse cross section as a function of photon virtuality. Results obtained with different wave functions are compared with the H1~\cite{Aktas:2005xu} and ZEUS~\cite{Chekanov:2004mw} data.
    }
    \label{fig:jpsi_R}
\end{figure}

To cancel normalization uncertainties, we next study cross section ratios. In Fig.~\ref{fig:jpsi_R} the longitudinal-to-transverse ratio of the \jpsi production cross section is shown as a function of the  photon virtuality. The results are compared with the H1 and ZEUS data from Refs.~\cite{Aktas:2005xu,Chekanov:2004mw}.  The first relativistic correction reduces the longitudinal cross section more than the transverse one. 
As discussed above, this is due to the fact that a part of the correction shifts the meson wave function away from the $\delta(z-1/2)$, which is the structure preferred by longitudinal photons but not by transverse photons.  This shows up as a decrease in the longitudinal to transverse ratio as a function of $Q^2$. The effect is even stronger with the Boosted Gaussian and BLFQ wavefunctions.

Finally, we study vector meson production in the future Electron Ion Collider. As the diffractive cross section at leading order in perturbative QCD is approximatively proportional to the squared gluon density, exclusive vector meson production is a promising observable to look for saturation effects at the future Electron Ion Collider (see e.g.~\cite{Mantysaari:2017slo}). 

To quantify the non-linear effects, we compute the nuclear suppression factor
\begin{equation}
    \frac{\sigma^{\gamma^* A \to \jpsim A}}{c A^{4/3} \sigma^{\gamma^* p \to \jpsim p}}.
\end{equation}
The denominator corresponds to the so called impulse approximation, which is used to transform the photon-proton cross section to the photon-nucleus case in the absence of nuclear effects, but taking into account the different form factors (transverse density profiles Fourier transformed to the momentum space).
The $A^{4/3}$ scaling can be understood to originate from the fact that the coherent cross section at $t=0$ scales as $\sim A^2$, and the width of the coherent spectra (location of the first diffractive minimum) is proportional to $1/R_A^2 \sim A^{-2/3}$. The numerical factor $c$ depends on the proton and nuclear form factors, and is found to be very close to $c=\frac{1}{2}$ in Ref.
~\cite{Mantysaari:2018nng}. In the absence of non-linear effects (or shadowing effects in the gluon distribution), with dipole amplitudes~\eqref{eq:ipsat_p} and \eqref{eq:ipsat_a} that depend linearly on $\rt^2 xg(x,\mu^2)$, this ratio is exactly $1$. 

\begin{figure}
    \includegraphics[width=0.5\textwidth]{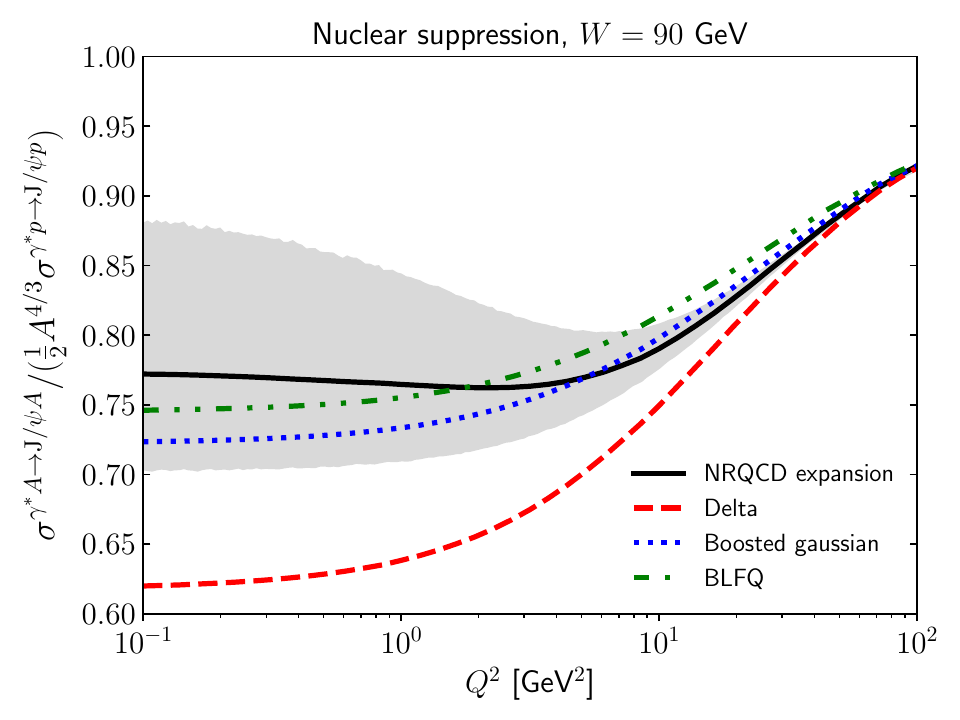}
    \caption{Nuclear suppression factor for total coherent \jpsi\ production as a function of $Q^2$ computed using the different vector meson wave functions. }
    \label{fig:jpsi_suppression}
\end{figure}

The obtained nuclear suppression factor is shown in Fig.~\ref{fig:jpsi_suppression} in the $Q^2$ range  accessible at the Electron Ion Collider. We emphasize that all the nuclear modifications in this figure are calculated with exactly the same dipole cross sections, corresponding to the same nuclear shadowing (as measured e.g. by the nuclear suppression in $F_L$ or $F_2$).  Thus the difference between the curves results purely from vector meson wave function effects.  When using the NRQCD wave function with the relativistic corretion, the Boosted Gaussian wave function or the BLFQ wave function, the obtained nuclear suppression factors are practically identical. Even though large mass of the vector meson renders the scale in the process large, a moderate suppression $\sim 0.75$ is found at small and moderate $Q
^2$. In the small $Q^2$ region the uncertainty obtained by varying the NRQCD matrix elements is large. 

The fully non relativistic wave function results in a much stronger suppression at small $Q^2$. This can be understood, as it was already seen in Fig.~\ref{fig:overlaps} that this wave function gives more weigh on larger dipoles compared to the other studied wave functions. As the larger dipoles are more sensitive to non-linear effects, a larger nuclear suppression in this case is anticipated. The first relativistic correction
$\sim -\rt^2$ suppresses the overlap at large dipole sizes, and consequently the nuclear suppression. At higher $Q^2$ the photon wave function again cuts out the large dipole part of the overlap in all cases, and as such the results obtained by applying the fully non-relativistic wave function do not differ from other wave functions any more. At asymptotically large $Q^2$ only small dipoles contribute and the dipole amplitudes can be linearized. Consequently, the suppression factor 
approaches unity at large $Q^2$ independently of the applied wave function.

The fact that the fully non-relativistic wave function results in a very different nuclear suppression demonstrates that the dependence on the meson wave function does not completely cancel in the nucleus-to-proton cross section ratios. Consequently, a  realistic (and relativistic) description of the vector meson wave function is necessary for interpreting the measured nuclear suppression factors. This indicates that there is a large theoretical uncertaintly in using the fully nonrelativistic formula of
Ryskin~\cite{Ryskin:1992ui}, not only for extracting absolute gluon distributions, but even for extracting  nuclear modifications to the  dipole cross section (or the gluon density) from cross section ratios.
\section{Conclusions}

In this work we proposed a new parametrization for the heavy vector  meson wave function based on NRQCD long-distance matrix elements. These matrix elements can be used to simultaneously constrain both the value and the derivative of the vector meson wave function at the origin using quarkonium decay data. This approach provides a systematic method to compute the vector meson wave function as an expansion in the strong coupling constant $\as$ and the quark velocity $v$. 

Compared to many phenomenological approaches used in the literature, our approach uses two independent constraints (the wave function value and its derivative). The obtained wave function is rotationally symmetric in the rest frame and contains only the $S$ wave component. Consequently, we simultaneously obtain a consistent parametrization for both polarization states. This is unlike in some widely used phenomenological parametrizations where the virtual photon like helicity structure is assumed on the light front. Relating light cone wavefunctions to rest frame ones also provides a consistent way to discuss the effect of a potential $D$-wave contribution to the meson wavefunction. We do not see indications, neither theoretically nor phenomenologically, that a significant $D$-wave contribution would be required or favored for the \jpsi\!.

The first relativistic correction to the wave function, controlled by the wave function derivative at the origin, is found to have a sizeable effect on the cross section. The negative $\sim -\rt^2$ relativistic contribution in terms of the transverse size $\rt$ suppresses the obtained wave function at larger dipole sizes. The momentum fraction part of the correction partially compensates for this effect for  the transverse photon by shifting the wave function away from the fully non relativistic configuration where both quarks carry the same fraction of the longitudinal momentum, a configuration which is not preferred by the transverse photon. 

A disadvantage in our approach is that it is not possible to obtain a wave function which is normalized to unity. In the NRQCD framework the value of the wavefunction at long distances is parametrized by a nonperturbative matrix element, whose effect is felt in the value of the wavefunction near the origin. This can lead to an overestimation of the cross section at $Q^2=0$, where one is most sensitive to the long distance behavior of the wave function. In practice, however, we obtain cross sections that are quite similar to what is given by e.g. the Boosted Gaussian parametrization. The wave function overlap with the photon is also smaller than with the BLFQ approach. Thus the lack of normalization in the wavefunction does not seem to be an important effect for \jpsi\!. The situation would be different for lighter vector mesons.

The structure of the wave function can be probed by studying cross sections (and cross section ratios) at different photon virtualities where the dipole sizes contributing to the cross section vary. The first relativistic correction is found to weaken the $Q^2$ dependence of the total \jpsi\ production cross section and the longitudinal-to-transverse ratio. These effects are broadly similar to predictions obtained by the Boosted Gaussian parametrization, or by the BLFQ wave function that is based on an explicit calculation on the light front including confinement effects.

When comparing vector meson production off protons to heavy nuclei, we find that the wave function does not completely cancel in the nuclear suppression factor, which compares the $\gamma^*A$ cross section to the $\gamma^*p$ in the  impulse approximation. This demonstrates that a realistic vector meson wave function is necessary to properly interpret the nuclear suppression results, and in particular a fully non-relativistic approach can not be reliably used to extract the non-linear effects on the nuclear structure.

In addition to the corrections in velocity, it would be important to include perturbative corrections in the strong coupling $\as$ in the calculation of exclusive vector meson production. Indeed some recent advances~\cite{Boussarie:2016bkq,Escobedo:2019bxn} are gradually making it possible to do so in the dipole picture. However, a study of the phenomenological implications of these $\as$ corrections remains to be done. In terms of understanding current and future experimental collider data, it would also be important to explore whether this approach can be extended to excited states such as the $\psi(nS)$.

\section*{Acknowledgements}
We thank M. Escobedo and M. Li for discussions.
This work was supported by the Academy of Finland, projects 314764 (H.M), 321840 (T.L and J.P) and 314162 (J.P).  T.L is supported by the European Research Council (ERC) under the European Union’s Horizon 2020 research and innovation programme (grant agreement No ERC-2015-CoG-681707). The content of this article does not reflect the official opinion of the European Union and responsibility for the information and views expressed therein lies entirely with the authors. %

\appendix

\section{Orbital decomposition}
\label{app:s_d}

In Sec.~\ref{sec:nrqcd} we highlighted how it is crucial to properly transform the NRQCD based vector meson wave function to the light front by performing the Melosh rotation.
In particular, we demonstrated that this rotation gives rise to the helicity structures absent in the rest frame spin structure (e.g. non-zero $\Psi^{\lambda=\pm 1}_{h=\pm 1, \bar h=\mp 1})$.

In this section, we illustrate the role of the Melosh rotation by considering both the  \jpsi and virtual photon (in the case of charm quarks) wave functions, and determining the contributions from the $S$ and $D$ wave components. The NRQCD based wave function obtained in Sec.~\ref{sec:nrqcd} contains only the $S$ wave structure. For the \jpsi wave function, we study here the commonly used Boosted Gaussian parametrization (see Sec.~\ref{sec:boosted_gaussian}).

The $S$ and $D$ waves are properly defined in the rest frame. Consequently, we take the vector meson or the virtual photon wave functions on the light front written in momentum space and perform the Melosh rotation to transform them to the meson rest frame. In the rest frame we then remove either the $S$ or $D$ wave contribution, and transform the final wave function back to the light front and Fourier transform to  transverse coordinate space.

\begin{figure}
    \includegraphics[width=0.5\textwidth]{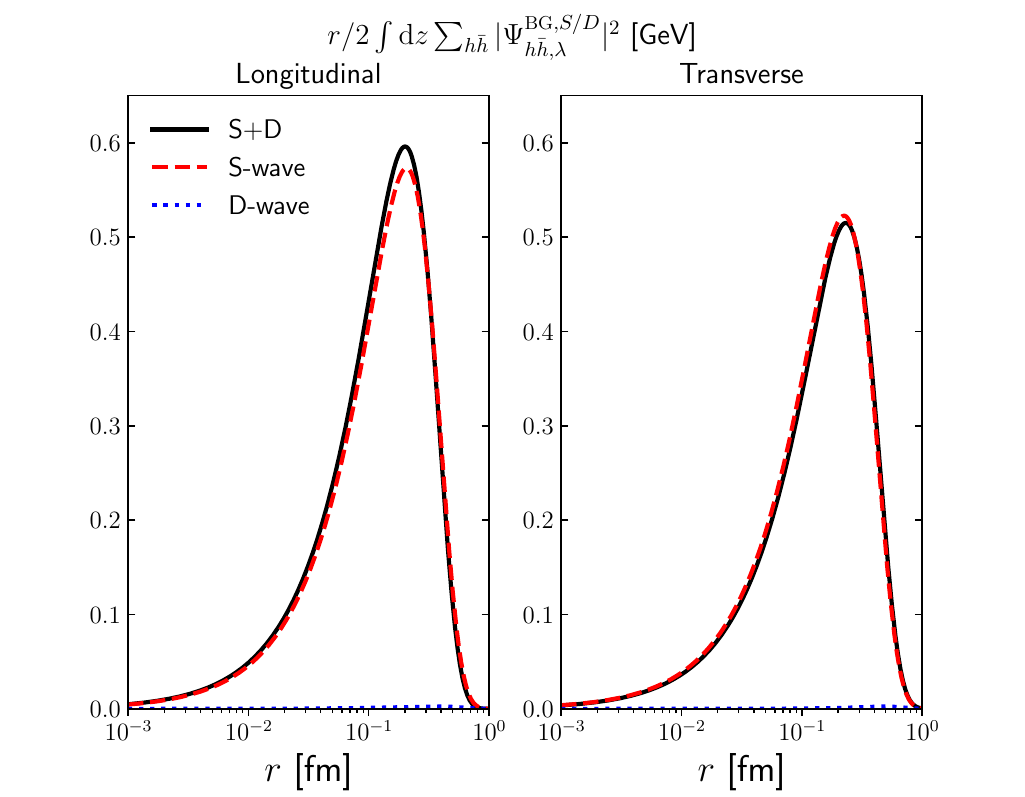}
    \caption{Vector meson wave function from the Boosted Gaussian parametrization decomposed into $S$ and $D$ wave components in the \jpsi rest frame as a function of the quark-antiquark transverse separation. Left panel shows the longitudinal polarization and right panel transverse polarization. 
   }
    \label{fig:jpsi_bg_s_d}
\end{figure}

The Boosted Gaussian parametrization of the \jpsi wave function is decomposed to $S$ and $D$ components in Fig.~\ref{fig:jpsi_bg_s_d}. In principle the angular momentum structure of the parametrization could turn out to correspond to a large $D$-wave component in the rest frame. Indeed it is mostly constrained by the choice of having helicity structure on the light front exactly the same as that of the photon, which has a large $D$-wave component as we will see. However, in practice the $S$-wave only result is a good approximation of the full result. This is due to the small quark velocities contributing to the wave function, as in the momentum space the Boosted Gaussian wave function is exponentially suppressed at large invariant mass $M^2=\frac{\kt^2+m_c^2}{z(1-z)}$. Thus, large transverse momentum $|\kt|$ or large longitudinal momentum ($z\to 0$ or $z\to 1$) contributions are heavily suppressed, and do not generate a significant $D$-wave component. 

A similar discussion can be carried out for  the BLFQ wave function described in Sec.~\ref{sec:boosted_gaussian}. As shown in~\cite{Li:2018uif}, 
in the rest frame, the squared \jpsi BLFQ wave function is dominated by the $S$ wave component, the $D$ wave contributing only a small fraction of the order of $0.1 \dots 4\%$ (depending on the polarization). In heavier mesons, this contribution is even smaller. This is comparable to the Boosted Gaussian case discussed above. 

Overall, based on neither the Boosted Gaussian nor the BLFQ parametrizations, we do not see any confirmation for the result of Ref.~\cite{Krelina:2019egg}, where the $D$ wave part of the \jpsi wave function was found to result in tens of percent contribution on the vector meson production cross section. Part of this discrepancy might be merely a question of terminology. In our discussion here, we have insisted that the terms $S$-wave and $D$-wave refer to the angular momentum components of the 3-dimensional wave function in the meson rest frame. Thus the mere presence, in the light cone wave function, of terms proportional to transverse momenta originating from the Melosh rotation cannot be taken as an indication of a $D$-wave component in the meson.

\begin{figure}
    \includegraphics[width=0.5\textwidth]{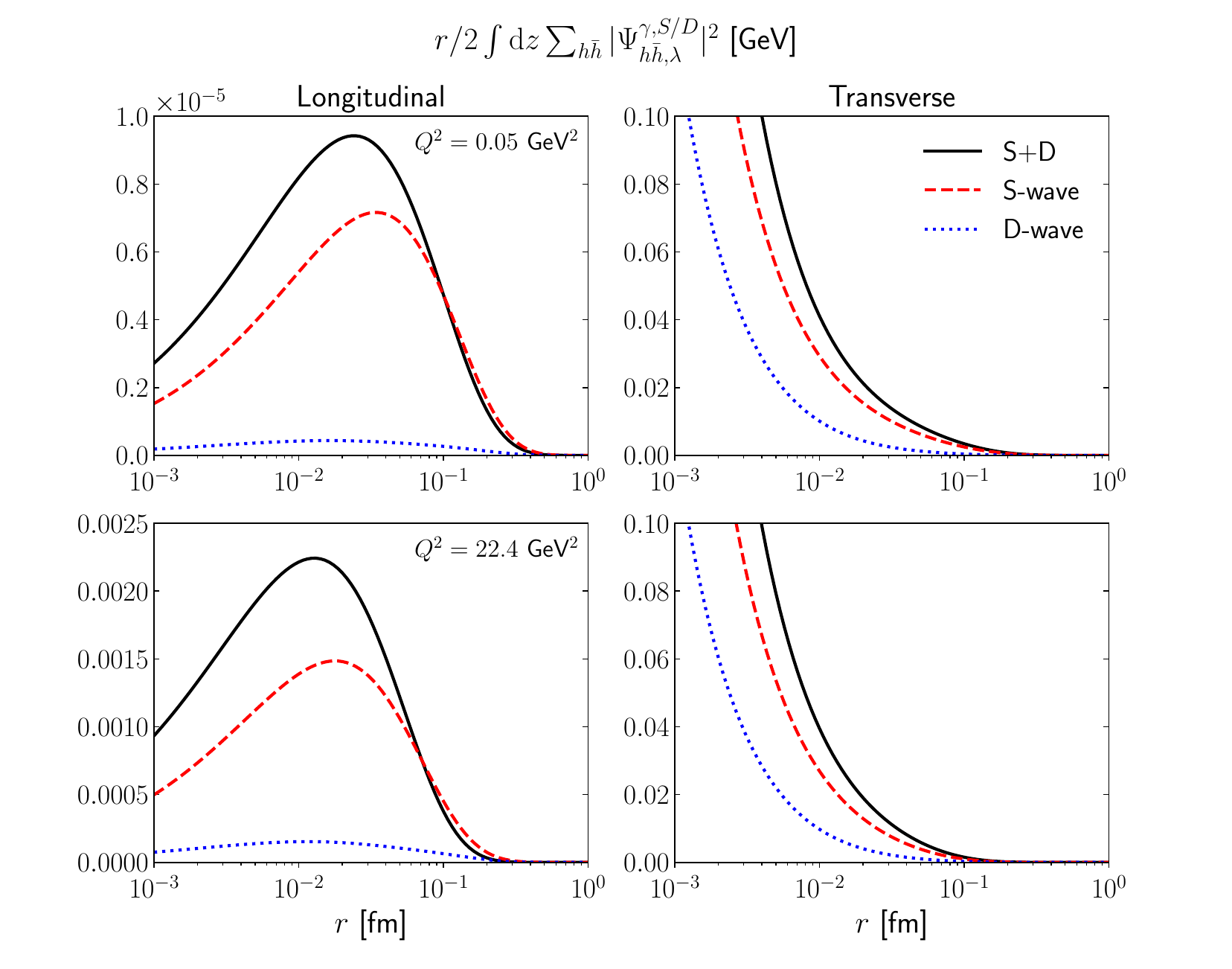}
    \caption{Virtual photon wave function integrated over the longitudinal momentum fraction $z$ decomposed to $S$ and $D$ wave components as a function of the quark-antiquark transverse separation.  }
    \label{fig:photon_s_d}
\end{figure}

Let us now move to the case of a virtual photon. Since a spacelike virtual photon does not have a rest frame and is not a bound state, it is not customarily thought of in terms of an $S$ -- $D$ wave decomposition. Now, however, we have an explicit light cone wave function for the photon just like for the meson, and we can use the same procedure to determine its $S$- and $D$-wave components in the meson rest frame.  The resulting squared light front wave functions summed over quark helicities are shown in Fig.~\ref{fig:photon_s_d}. The full photon wave function, written in Eqs.~\eqref{eq:photon_L} and \eqref{eq:photon_T}, is denoted by $S+D$, as it can be written as a sum of these two components. When compared to the full result, the squared $D$ wave only contribution is found to be strongly suppressed. There is also a contribution originating from the overlap between the $S$ and $D$ wave contributions. This term would vanish if we integrated over all the angles. Here, we only integrate over the azimuthal direction of $\rt$. Integration over the momentum fraction $z$ corresponds to the evaluation of the coordinate space wave function at $x^3=0$, and consequently one angular integral is not performed and the overlap does not vanish. The relative importance of different contributions is found to be approximatively independent of $Q^2$.

The $S-D$ overlap contribution is numerically significant, which is reflected by the large difference between the full result and the $S$ wave only contribution. This suggests that even though the $D$ wave contribution is suppressed by the quark velocity, the charm quark mass is not large enough to render this contribution negligible. This is due to the fact that the photon wave function in the momentum space behaves as $\sim 1/\kt^2$ where $\kt$ is the quark transverse momentum, and this powerlike tail brings numerically large contributions from relatively large momenta. Additionally, the integration over the longitudinal momentum fraction $z$ includes high momentum contributions, as the photon wave function has a support over a large range of $z$.

The contribution from the $S-D$ overlap changes sign at large transverse separations in case of the longitudinal polarization. 
There is no node in the radial part of the wave function, but the spherical harmonic function describing the angular part of the $D$ wave component changes sign, which explains the sign flip. 
In the $S-D$ overlap mostly 
the  helicity-$+-$ and $-+$ components of the $D$ wave contribute by coupling to the $S$ wave. On the other hand, in the $D$ wave squared wave function, one sums all helicity components. As the $D$ wave component itself is a relativistic correction, none of the helicity structures dominates unlike in the $S$ wave part. Moreover, only the $+-$ and $-+$ helicity components change sign at large distances, and as the $++$ and $--$ components do not vanish in this region, no node appears in the squared $D$ wave result. In the case of transverse polarization with $\lambda=\pm 1$, the helicity component $\pm \pm$ in the $D$ wave also changes the sign at large distances, but this effect is not easily visible in Fig.~\ref{fig:photon_s_d} as the other helicity components that do not change sign dominate.

\section{Photon-like parametrizations of light cone wave functions}
\label{app:scalarlcwf}

As discussed in Sec.~\ref{sec:pheno-wf}, an often used approach to parametrize  vector meson wave functions is to start from the helicity structure of the virtual photon light cone wave functions
\eqref{eq:photon_L},  \eqref{eq:photon_T}. One then 
replaces the Bessel function $K_0$ in  the photon wave functions \eqref{eq:photon_L} and \eqref{eq:photon_T} 
by an unknown function as~\cite{Kowalski:2006hc}
\begin{equation}
    e_f e z(1-z) \frac{K_0(\epsilon r)}{2\pi} \to \phi_{T,L}(r,z),
\end{equation}
with the explicit factor $Q$ in the longitudinal wave function  replaced by the meson mass as $2Q \to M_V$.
This leads, with our sign conventions,  to the wave function being written as
\begin{multline}   
    \psi^{\lambda=0}_{h \bar h} (\rt, z) 
    = \sqrt{\nc} \delta_{h, -\bar h} 
    \\ \times
    \left[M_V+ \frac{m_c^2-\nabla^2_\rt}{M_V z (1-z)} \right] \phi_L(\rt, z)
\label{eq:kmwformL}
\end{multline} 
\begin{multline}   
\psi^{\lambda=\pm 1}_{h \bar h} (\rt, z) = \sqrt{2\nc} \frac{1}{z(1-z)} \left( m_c \delta_{h, \pm}\delta_{\bar h, \pm} \right. 
 \\ 
    \left. \mp i e^{\pm i\theta_r} \left(z \delta_{h, \pm} \delta_{\bar h, \mp} -(1-z)\delta_{h, \mp} \delta_{\bar h, \pm} \right) \partial_r \right)\phi_T(\rt, z).
\label{eq:kmwformT}
\end{multline}

The scalar functions $\phi_{T,L}(\rt,z)$ are then parametrized, and the parameters can be determined by requiring that the resulting wave function is normalized to unity and reproduces the experimental leptonic decay width. In terms of Lorentz-invariant form factors this means that the meson is assumed to  have a nonzero Dirac form factor but a vanishing Pauli form factor, since this is the structure dictated by the gauge-boson-fermion vertex at leading order perturbation theory. The procedure therefore does not generate the most general possible helicity structure.

This photon-like parametrization approach starts from a spacelike photon, where the photon momentum breaks rotational symmetry that is manifested here as a symmetry between longitudinal and transverse meson polarization states. The common approach is to separately parametrize the longitudinal and transverse functions $\phi_{T,L}(\rt,z)$. The helicity structure obtained by generalization from the photon wave function is of course consistent with rotational symmetry, since the decay of a timelike virtual photon is rotationally symmetric. Thus one could derive a constraint relating $\phi_{L}(\rt,z)$ and $\phi_{T}(\rt,z)$ by requiring the meson rest frame wavefunctions to be the same. To our knowledge this approach has not, however, been used in the literature. Using separate parametrizations for  $\phi_{T,L}(\rt,z)$ should be contrasted with the approach in this paper. Here, we maintain rotational invariance in the meson rest frame, in particular starting from the same decay constants calculated from the rest frame wave functions. Our procedure for going from the rest frame to the light cone wave function therefore simultaneously determines the wavefunction for both longitudinal and transverse polarization states.

One can take the parametrization \eqref{eq:kmwformL}, \eqref{eq:kmwformT} in momentum space, perform the inverse Melosh rotation and separate the $S$- and $D$- wave components to get a rest frame 3-dimensional wavefunction. Assuming that the Fourier transforms of the scalar functions are rotationally invariant, i.e. $\phi_{T, L}(\kt, z)$ depend only on $k = \lvert \vec k \rvert = \sqrt{(M/2)^2-m_c^2}$, the result of this exercise in momentum space is:
\begin{multline}
    \psi_S^{\lambda=0} = \phi_L(k)\left(M_V+\frac{4E^2}{M_V}\right)\\
    \times \sqrt{N_c \pi} \frac{2}{(E+m_c)(2E)^{3/2}} \left(\frac{1}{3}k^2+(E+m)^2\right)
\end{multline}
\begin{multline}
    \psi_S^{\lambda=+1,-1} = \phi_T(k) \cdot 4E \\
   \times \sqrt{N_c \pi} \frac{2}{(E+m_c)(2E)^{3/2}} \left(\frac{1}{3}k^2+(E+m)^2\right)
\end{multline}
\begin{multline}
\label{eq:L_D}
    \psi_D^{\lambda=0} = \phi_L(k)\left(M_V+\frac{4E^2}{M_V}\right)\\
    \times \sqrt{N_c \pi} \frac{2}{(E+m_c)(2E)^{3/2}} \frac{4}{3\sqrt{2}} k^2
\end{multline}
\begin{multline}
\label{eq:T_D}
    \psi_D^{\lambda=+1,-1} = \phi_T(k)\cdot 4E \\
    \times \sqrt{N_c \pi} \frac{2}{(E+m_c)(2E)^{3/2}}\frac{4}{3\sqrt{2}} k^2 .
\end{multline}
These expressions are written in terms of the energy of the quark in the meson rest frame $E=\sqrt{\vec k^2 + m_c^2} = M/2$, and $k=|\vec k|$. Let us point out a few aspects of these expressions. Firstly, as discussed above, in the photon-like parametrization there is always a $D$-wave component in the meson wave function.
It is, as expected, explicitly a relativistic correction, i.e. proportional to the squared 3-momentum of the quark. Secondly, the rest frame wave functions of the transverse and longitudinal polarizations are not the same, but differ by a factor $\left(M_V+\frac{4E^2}{M_V}\right)/(4E) \approx 1 + \mathcal{O}\left((M-M_V)^2\right)$, where $M$ is the invariant mass of the quark pair and $M_V$ the mass of the meson. As discussed  earlier in Sec.~\ref{sec:nrqcd}, the coordinate transformation from $k^3$ to $z$ inevitably introduces ambiguities that are proportional to this difference, so this should not come as a surprise. In an NRQCD power counting, this difference is of the order of the binding energy of the meson, which is higher order than we are considering here.

In spite of this discussion, the wavefunction that we obtained  in Sec.~\ref{sec:pheno-wf} can in fact be written in a photon-like form in terms of scalar parts of light cone wave functions. In the notation of \cite{Kowalski:2006hc} these read 
\begin{multline}\label{eq:nrqcdphiL}
    \phi_L(\rt, z) =  \frac{\pi}{\sqrt{N_c}(2m_{c,NR})^{3/2}} \frac{4M_V m_{c,NR}}{4m_{c,NR}^2+M_V^2} \\
    \cdot \Bigg[A \delta(z-1/2)  \\
     +\frac{B}{m^2_{c,NR}} \Bigg( \left(\frac{34m_{c,NR}^2+\frac{5}{2} M_V^2}{4m_{c,NR}^2+M_V^2}+m^2_{c,NR}\rt^2 \right) \delta(z-1/2)  \\
     - \frac{1}{4}\partial_z^2 \delta(z-1/2) \Bigg) \Bigg]
\end{multline}
\begin{multline}\label{eq:nrqcdphiT}
    \phi_T(\rt, z) = \frac{\pi}{\sqrt{N_c}(2m_{c,NR})^{3/2}} \Bigg[A \delta(z-1/2)  \\
     +\frac{B}{m^2_{c,NR}} \Bigg( \left(\frac{11}{2}+m^2_{c,NR}\rt^2\right) \delta(z-1/2)  \\
     -\frac{1}{4}\partial_z^2 \delta(z-1/2)  \Bigg) \Bigg]
\end{multline}
We emphasize that we do not expect that writing down such a paramterization in terms of two scalar parts of a light cone wave function having the helicity structure of a photon would be possible at higher orders in the nonrelativistic expansion.

As a side remark, we discussed above that the photon-like structure generically implies a non-zero $D$ wave component, see \eqs\eqref{eq:L_D} and~\eqref{eq:T_D}. On the other hand, our NRQCD based wave function by construction has no $D$ component. However, the $D$-wave component resulting from inserting the scalar parts~\eqref{eq:nrqcdphiL} and~\eqref{eq:nrqcdphiT} into the formulae for the $D$-wave contribution, \eqs\eqref{eq:L_D} and~\eqref{eq:T_D}, behaves as $\sim k^2 \nabla_{\vec{k}}^2 \delta^{(3)}(\vec{k})$. Such a function actually yields zero when convoluted with any test function  $f(\vec{k})$, since the angular integral picks out the $\ell=2$  component of $f$, which must vanish at $k=0$. Thus the $D$-wave contribution corresponding to \eqref{eq:nrqcdphiL} and~\eqref{eq:nrqcdphiT} is in fact zero in a distribution sense. 

It is interesting to note, that when one calculates from these expressions the decay constants for the different polarization states using the light cone perturbation theory expressions~(26) and (27) in Ref.~\cite{Kowalski:2006hc}, one obtains:
\begin{eqnarray}
\label{eq:ftkmwcompL}
f_L &=& \sqrt{\frac{2 \nc}{\mcnr}} e_f \left(A+\frac{5}{2} \frac{B}{\mcnr^2} \right)
\\
\label{eq:ftkmwcompT}
f_T &=& \sqrt{\frac{2 \nc}{\mcnr}} e_f \frac{2\mcnr}{M_V}\left(A-\frac{1}{2}\frac{B}{\mcnr^2}\right)
\end{eqnarray}
The results are not exactly equal. However, as discussed above, if one approximates the meson mass by  the quark pair invariant mass, as we did in transforming to the momentum fraction $z$, they do reduce to the same result. This can be seen explicitly by replacing $M_V$ in \eqref{eq:ftkmwcompT} by $\langle M \rangle \approx 2 \sqrt{\mcnr^2 + \langle \mathbf{q}^2\rangle }$ and using \eq\eqref{eq:B_coefficient} to write, at lowest nontrivial order in the quark velocity,  $2\mcnr/M_V \approx 1 + 3B/(\mcnr^2 A)$ (note that $B<0$).  In this approximation \eqs\eqref{eq:ftkmwcompL} and~\eqref{eq:ftkmwcompT} also give back the same decay width expression that we are using to determine the rest frame wavefunction.  We reiterate that a difference such as this can be expected in our procedure. We are constructing our wavefunctions by requiring the decay widths calculated from the rest frame wave functions to have the correct value, and to be the same for the different polarization states. The coordinate transformation to light cone wave functions does not conserve these properties exactly, but only  up to a given order in the nonrelativistic expansion.

\bibliographystyle{JHEP-2modlong.bst}
\bibliography{refs}

\providecommand{\href}[2]{#2}\begingroup\raggedright\begin{thebibliography}{100}

\bibitem{Gelis:2010nm}
F.~Gelis, E.~Iancu, J.~Jalilian-Marian and R.~Venugopalan, {\it {The Color
  Glass Condensate}},
  \href{http://dx.doi.org/10.1146/annurev.nucl.010909.083629}{{\em Ann. Rev.
  Nucl. Part. Sci.} {\bf 60} (2010) 463}
  [\href{http://arXiv.org/abs/1002.0333}{{\tt arXiv:1002.0333 [hep-ph]}}].

\bibitem{Albacete:2014fwa}
J.~L. Albacete and C.~Marquet, {\it {Gluon saturation and initial conditions
  for relativistic heavy ion collisions}},
  \href{http://dx.doi.org/10.1016/j.ppnp.2014.01.004}{{\em Prog. Part. Nucl.
  Phys.} {\bf 76} (2014) 1} [\href{http://arXiv.org/abs/1401.4866}{{\tt
  arXiv:1401.4866 [hep-ph]}}].

\bibitem{Ryskin:1992ui}
M.~G. Ryskin, {\it {Diffractive $\mathrm{J}/\psi$ electroproduction in LLA
  QCD}},  \href{http://dx.doi.org/10.1007/BF01555742}{{\em Z. Phys.} {\bf C57}
  (1993) 89}.

\bibitem{Aktas:2005xu}
{\bf H1} collaboration, A.~Aktas {\em et.~al.}, {\it {Elastic $\mathrm{J}/\psi$
  production at HERA}},
  \href{http://dx.doi.org/10.1140/epjc/s2006-02519-5}{{\em Eur. Phys. J.} {\bf
  C46} (2006) 585} [\href{http://arXiv.org/abs/hep-ex/0510016}{{\tt
  arXiv:hep-ex/0510016 [hep-ex]}}].

\bibitem{Chekanov:2002xi}
{\bf ZEUS} collaboration, S.~Chekanov {\em et.~al.}, {\it {Exclusive
  photoproduction of $\mathrm{J}/\psi$ mesons at HERA}},
  \href{http://dx.doi.org/10.1007/s10052-002-0953-7}{{\em Eur. Phys. J.} {\bf
  C24} (2002) 345} [\href{http://arXiv.org/abs/hep-ex/0201043}{{\tt
  arXiv:hep-ex/0201043 [hep-ex]}}].

\bibitem{Chekanov:2002rm}
{\bf ZEUS} collaboration, S.~Chekanov {\em et.~al.}, {\it {Measurement of
  proton dissociative diffractive photoproduction of vector mesons at large
  momentum transfer at HERA}},
  \href{http://dx.doi.org/10.1140/epjc/s2002-01079-0}{{\em Eur. Phys. J.} {\bf
  C26} (2003) 389} [\href{http://arXiv.org/abs/hep-ex/0205081}{{\tt
  arXiv:hep-ex/0205081 [hep-ex]}}].

\bibitem{Aktas:2003zi}
{\bf H1} collaboration, A.~Aktas {\em et.~al.}, {\it {Diffractive
  photoproduction of $\mathrm{J}/\psi$ mesons with large momentum transfer at
  HERA}},  \href{http://dx.doi.org/10.1016/j.physletb.2003.06.056}{{\em Phys.
  Lett.} {\bf B568} (2003) 205}
  [\href{http://arXiv.org/abs/hep-ex/0306013}{{\tt arXiv:hep-ex/0306013
  [hep-ex]}}].

\bibitem{Chekanov:2004mw}
{\bf ZEUS} collaboration, S.~Chekanov {\em et.~al.}, {\it {Exclusive
  electroproduction of $\mathrm{J}/\psi$ mesons at HERA}},
  \href{http://dx.doi.org/10.1016/j.nuclphysb.2004.06.034}{{\em Nucl. Phys.}
  {\bf B695} (2004) 3} [\href{http://arXiv.org/abs/hep-ex/0404008}{{\tt
  arXiv:hep-ex/0404008 [hep-ex]}}].

\bibitem{Alexa:2013xxa}
{\bf H1} collaboration, C.~Alexa {\em et.~al.}, {\it {Elastic and
  Proton-Dissociative Photoproduction of $\mathrm{J}/\psi$ Mesons at HERA}},
  \href{http://dx.doi.org/10.1140/epjc/s10052-013-2466-y}{{\em Eur. Phys. J.}
  {\bf C73} (2013)~no.~6 2466} [\href{http://arXiv.org/abs/1304.5162}{{\tt
  arXiv:1304.5162 [hep-ex]}}].

\bibitem{Adloff:1999kg}
{\bf H1} collaboration, C.~Adloff {\em et.~al.}, {\it {Elastic
  electroproduction of $\rho$ mesons at HERA}},
  \href{http://dx.doi.org/10.1007/s100520050703}{{\em Eur. Phys. J.} {\bf C13}
  (2000) 371} [\href{http://arXiv.org/abs/hep-ex/9902019}{{\tt
  arXiv:hep-ex/9902019 [hep-ex]}}].

\bibitem{Chekanov:2005cqa}
{\bf ZEUS} collaboration, S.~Chekanov {\em et.~al.}, {\it {Exclusive
  electroproduction of $\phi$ mesons at HERA}},
  \href{http://dx.doi.org/10.1016/j.nuclphysb.2005.04.009}{{\em Nucl. Phys.}
  {\bf B718} (2005) 3} [\href{http://arXiv.org/abs/hep-ex/0504010}{{\tt
  arXiv:hep-ex/0504010 [hep-ex]}}].

\bibitem{Aaron:2009xp}
{\bf H1} collaboration, F.~D. Aaron {\em et.~al.}, {\it {Diffractive
  Electroproduction of $\rho$ and $\phi$ Mesons at HERA}},
  \href{http://dx.doi.org/10.1007/JHEP05(2010)032}{{\em JHEP} {\bf 05} (2010)
  032} [\href{http://arXiv.org/abs/0910.5831}{{\tt arXiv:0910.5831 [hep-ex]}}].

\bibitem{Chekanov:2009zz}
{\bf ZEUS} collaboration, S.~Chekanov {\em et.~al.}, {\it {Exclusive
  photoproduction of $\Upsilon$ mesons at HERA}},
  \href{http://dx.doi.org/10.1016/j.physletb.2009.07.066}{{\em Phys. Lett.}
  {\bf B680} (2009) 4} [\href{http://arXiv.org/abs/0903.4205}{{\tt
  arXiv:0903.4205 [hep-ex]}}].

\bibitem{Adloff:2000vm}
{\bf H1} collaboration, C.~Adloff {\em et.~al.}, {\it {Elastic photoproduction
  of $\mathrm{J}/\psi$ and Upsilon mesons at HERA}},
  \href{http://dx.doi.org/10.1016/S0370-2693(00)00530-X}{{\em Phys. Lett.} {\bf
  B483} (2000) 23} [\href{http://arXiv.org/abs/hep-ex/0003020}{{\tt
  arXiv:hep-ex/0003020 [hep-ex]}}].

\bibitem{Bertulani:2005ru}
C.~A. Bertulani, S.~R. Klein and J.~Nystrand, {\it {Physics of ultra-peripheral
  nuclear collisions}},
  \href{http://dx.doi.org/10.1146/annurev.nucl.55.090704.151526}{{\em Ann. Rev.
  Nucl. Part. Sci.} {\bf 55} (2005) 271}
  [\href{http://arXiv.org/abs/nucl-ex/0502005}{{\tt arXiv:nucl-ex/0502005
  [nucl-ex]}}].

\bibitem{Klein:2019qfb}
S.~R. Klein and H.~Mäntysaari, {\it {Imaging the nucleus with high-energy
  photons}},  \href{http://dx.doi.org/10.1038/s42254-019-0107-6}{{\em Nature
  Rev. Phys.} {\bf 1} (2019)~no.~11 662}
  [\href{http://arXiv.org/abs/1910.10858}{{\tt arXiv:1910.10858 [hep-ex]}}].

\bibitem{Afanasiev:2009hy}
{\bf PHENIX} collaboration, S.~Afanasiev {\em et.~al.}, {\it {Photoproduction
  of $\mathrm{J}/\psi$ and of high mass $e^+e^-$ in ultra-peripheral Au+Au
  collisions at $\sqrt{s} = 200$ GeV}},
  \href{http://dx.doi.org/10.1016/j.physletb.2009.07.061}{{\em Phys. Lett.}
  {\bf B679} (2009) 321} [\href{http://arXiv.org/abs/0903.2041}{{\tt
  arXiv:0903.2041 [nucl-ex]}}].

\bibitem{TheALICE:2014dwa}
{\bf ALICE} collaboration, B.~B. Abelev {\em et.~al.}, {\it {Exclusive
  $\mathrm{J/}\psi$ photoproduction off protons in ultra-peripheral p-Pb
  collisions at $\sqrt{s_{\rm NN}}=5.02$ TeV}},
  \href{http://dx.doi.org/10.1103/PhysRevLett.113.232504}{{\em Phys. Rev.
  Lett.} {\bf 113} (2014)~no.~23 232504}
  [\href{http://arXiv.org/abs/1406.7819}{{\tt arXiv:1406.7819 [nucl-ex]}}].

\bibitem{Acharya:2018jua}
{\bf ALICE} collaboration, S.~Acharya {\em et.~al.}, {\it {Energy dependence of
  exclusive $\mathrm {J}/\psi $ photoproduction off protons in ultra-peripheral
  p–Pb collisions at $\sqrt{s_{\mathrm {\scriptscriptstyle NN}}} = 5.02$
  TeV}},  \href{http://dx.doi.org/10.1140/epjc/s10052-019-6816-2}{{\em Eur.
  Phys. J.} {\bf C79} (2019)~no.~5 402}
  [\href{http://arXiv.org/abs/1809.03235}{{\tt arXiv:1809.03235 [nucl-ex]}}].

\bibitem{Abbas:2013oua}
{\bf ALICE} collaboration, E.~Abbas {\em et.~al.}, {\it {Charmonium and
  $e^+e^-$ pair photoproduction at mid-rapidity in ultra-peripheral Pb-Pb
  collisions at $\sqrt{s_{\rm NN}}$=2.76 TeV}},
  \href{http://dx.doi.org/10.1140/epjc/s10052-013-2617-1}{{\em Eur. Phys. J.}
  {\bf C73} (2013)~no.~11 2617} [\href{http://arXiv.org/abs/1305.1467}{{\tt
  arXiv:1305.1467 [nucl-ex]}}].

\bibitem{Abelev:2012ba}
{\bf ALICE} collaboration, B.~Abelev {\em et.~al.}, {\it {Coherent $J/\psi$
  photoproduction in ultra-peripheral Pb-Pb collisions at $\sqrt{s_{NN}} =
  2.76$ TeV}},  \href{http://dx.doi.org/10.1016/j.physletb.2012.11.059}{{\em
  Phys. Lett.} {\bf B718} (2013) 1273}
  [\href{http://arXiv.org/abs/1209.3715}{{\tt arXiv:1209.3715 [nucl-ex]}}].

\bibitem{Khachatryan:2016qhq}
{\bf CMS} collaboration, V.~Khachatryan {\em et.~al.}, {\it {Coherent $J/\psi$
  photoproduction in ultra-peripheral PbPb collisions at $\sqrt {s_{NN}} =$
  2.76 TeV with the CMS experiment}},
  \href{http://dx.doi.org/10.1016/j.physletb.2017.07.001}{{\em Phys. Lett.}
  {\bf B772} (2017) 489} [\href{http://arXiv.org/abs/1605.06966}{{\tt
  arXiv:1605.06966 [nucl-ex]}}].

\bibitem{Sirunyan:2018sav}
{\bf CMS} collaboration, A.~M. Sirunyan {\em et.~al.}, {\it {Measurement of
  exclusive $\Upsilon$ photoproduction from protons in pPb collisions at
  $\sqrt{s_\mathrm{NN}} =$ 5.02 TeV}},
  \href{http://dx.doi.org/10.1140/epjc/s10052-019-6774-8}{{\em Eur. Phys. J.}
  {\bf C79} (2019)~no.~3 277} [\href{http://arXiv.org/abs/1809.11080}{{\tt
  arXiv:1809.11080 [hep-ex]}}].

\bibitem{Adam:2019rxb}
{\bf STAR} collaboration, J.~Adam, {\it {Coherent $J/\psi$ photoproduction in
  ultra-peripheral collisions at STAR}},
  \href{http://dx.doi.org/10.22323/1.352.0042}{{\em PoS} {\bf DIS2019} (2019)
  042}.

\bibitem{LHCb:2018ofh}
{\bf LHCb} collaboration, A.~Bursche, {\it {Study of coherent $J/\psi$
  production in lead-lead collisions at $\sqrt{s_{\rm NN}} =5\ \rm{TeV}$ with
  the LHCb experiment}},
  \href{http://dx.doi.org/10.1016/j.nuclphysa.2018.10.069}{{\em Nucl. Phys.}
  {\bf A982} (2019) 247}.

\bibitem{Li:2020ntd}
{\bf LHCb} collaboration, H.~Li in {\em {28th International Conference on
  Ultrarelativistic Nucleus-Nucleus Collisions (Quark Matter 2019) Wuhan,
  China, November 4-9, 2019}}, 2020.
\newblock \href{http://arXiv.org/abs/2002.01863}{{\tt arXiv:2002.01863
  [nucl-ex]}}.

\bibitem{Lappi:2013am}
T.~Lappi and H.~Mantysaari, {\it {$\mathrm{J}/\psi$ production in
  ultraperipheral Pb+Pb and $p$+Pb collisions at energies available at the CERN
  Large Hadron Collider}},
  \href{http://dx.doi.org/10.1103/PhysRevC.87.032201}{{\em Phys. Rev.} {\bf
  C87} (2013)~no.~3 032201} [\href{http://arXiv.org/abs/1301.4095}{{\tt
  arXiv:1301.4095 [hep-ph]}}].

\bibitem{Guzey:2013qza}
V.~Guzey and M.~Zhalov, {\it {Exclusive $J/{\psi}$ production in
  ultraperipheral collisions at the LHC: constrains on the gluon distributions
  in the proton and nuclei}},
  \href{http://dx.doi.org/10.1007/JHEP10(2013)207}{{\em JHEP} {\bf 10} (2013)
  207} [\href{http://arXiv.org/abs/1307.4526}{{\tt arXiv:1307.4526 [hep-ph]}}].

\bibitem{Guzey:2016piu}
V.~Guzey, E.~Kryshen and M.~Zhalov, {\it {Coherent photoproduction of vector
  mesons in ultraperipheral heavy ion collisions: Update for run 2 at the CERN
  Large Hadron Collider}},
  \href{http://dx.doi.org/10.1103/PhysRevC.93.055206}{{\em Phys. Rev.} {\bf
  C93} (2016)~no.~5 055206} [\href{http://arXiv.org/abs/1602.01456}{{\tt
  arXiv:1602.01456 [nucl-th]}}].

\bibitem{Albacete:2010bs}
J.~L. Albacete and C.~Marquet, {\it {Single Inclusive Hadron Production at RHIC
  and the LHC from the Color Glass Condensate}},
  \href{http://dx.doi.org/10.1016/j.physletb.2010.02.073}{{\em Phys. Lett.}
  {\bf B687} (2010) 174} [\href{http://arXiv.org/abs/1001.1378}{{\tt
  arXiv:1001.1378 [hep-ph]}}].

\bibitem{Albacete:2012xq}
J.~L. Albacete, A.~Dumitru, H.~Fujii and Y.~Nara, {\it {CGC predictions for p +
  Pb collisions at the LHC}},
  \href{http://dx.doi.org/10.1016/j.nuclphysa.2012.09.012}{{\em Nucl. Phys.}
  {\bf A897} (2013) 1} [\href{http://arXiv.org/abs/1209.2001}{{\tt
  arXiv:1209.2001 [hep-ph]}}].

\bibitem{Lappi:2013zma}
T.~Lappi and H.~Mäntysaari, {\it {Single inclusive particle production at high
  energy from HERA data to proton-nucleus collisions}},
  \href{http://dx.doi.org/10.1103/PhysRevD.88.114020}{{\em Phys. Rev.} {\bf
  D88} (2013) 114020} [\href{http://arXiv.org/abs/1309.6963}{{\tt
  arXiv:1309.6963 [hep-ph]}}].

\bibitem{Ducloue:2015gfa}
B.~Ducloué, T.~Lappi and H.~Mäntysaari, {\it {Forward $J/\psi$ production in
  proton-nucleus collisions at high energy}},
  \href{http://dx.doi.org/10.1103/PhysRevD.91.114005}{{\em Phys. Rev.} {\bf
  D91} (2015)~no.~11 114005} [\href{http://arXiv.org/abs/1503.02789}{{\tt
  arXiv:1503.02789 [hep-ph]}}].

\bibitem{Ducloue:2016pqr}
B.~Ducloué, T.~Lappi and H.~Mäntysaari, {\it {Forward $J/\psi$ production at
  high energy: centrality dependence and mean transverse momentum}},
  \href{http://dx.doi.org/10.1103/PhysRevD.94.074031}{{\em Phys. Rev.} {\bf
  D94} (2016)~no.~7 074031} [\href{http://arXiv.org/abs/1605.05680}{{\tt
  arXiv:1605.05680 [hep-ph]}}].

\bibitem{Mantysaari:2019nnt}
H.~Mäntysaari and H.~Paukkunen, {\it {Saturation and forward jets in
  proton-lead collisions at the LHC}},
  \href{http://dx.doi.org/10.1103/PhysRevD.100.114029}{{\em Phys. Rev.} {\bf
  D100} (2019)~no.~11 114029} [\href{http://arXiv.org/abs/1910.13116}{{\tt
  arXiv:1910.13116 [hep-ph]}}].

\bibitem{Brodsky:1994kf}
S.~J. Brodsky, L.~Frankfurt, J.~F. Gunion, A.~H. Mueller and M.~Strikman, {\it
  {Diffractive leptoproduction of vector mesons in QCD}},
  \href{http://dx.doi.org/10.1103/PhysRevD.50.3134}{{\em Phys. Rev.} {\bf D50}
  (1994) 3134} [\href{http://arXiv.org/abs/hep-ph/9402283}{{\tt
  arXiv:hep-ph/9402283 [hep-ph]}}].

\bibitem{Anand:2018zle}
S.~Anand and T.~Toll, {\it {Exclusive diffractive vector meson production: A
  comparison between the dipole model and the leading twist shadowing
  approach}},  \href{http://dx.doi.org/10.1103/PhysRevC.100.024901}{{\em Phys.
  Rev.} {\bf C100} (2019)~no.~2 024901}
  [\href{http://arXiv.org/abs/1807.10888}{{\tt arXiv:1807.10888 [hep-ph]}}].

\bibitem{Hoodbhoy:1996zg}
P.~Hoodbhoy, {\it {Wave function corrections and off forward gluon
  distributions in diffractive $\mathrm{J}/\psi$ electroproduction}},
  \href{http://dx.doi.org/10.1103/PhysRevD.56.388}{{\em Phys. Rev.} {\bf D56}
  (1997) 388} [\href{http://arXiv.org/abs/hep-ph/9611207}{{\tt
  arXiv:hep-ph/9611207 [hep-ph]}}].

\bibitem{Frankfurt:1997fj}
L.~Frankfurt, W.~Koepf and M.~Strikman, {\it {Diffractive heavy quarkonium
  photoproduction and electroproduction in QCD}},
  \href{http://dx.doi.org/10.1103/PhysRevD.57.512}{{\em Phys. Rev.} {\bf D57}
  (1998) 512} [\href{http://arXiv.org/abs/hep-ph/9702216}{{\tt
  arXiv:hep-ph/9702216 [hep-ph]}}].

\bibitem{Mantysaari:2017dwh}
H.~Mäntysaari and B.~Schenke, {\it {Probing subnucleon scale fluctuations in
  ultraperipheral heavy ion collisions}},
  \href{http://dx.doi.org/10.1016/j.physletb.2017.07.063}{{\em Phys. Lett.}
  {\bf B772} (2017) 832} [\href{http://arXiv.org/abs/1703.09256}{{\tt
  arXiv:1703.09256 [hep-ph]}}].

\bibitem{Lappi:2010dd}
T.~Lappi and H.~Mantysaari, {\it {Incoherent diffractive
  $\mathrm{J}/\psi$-production in high energy nuclear DIS}},
  \href{http://dx.doi.org/10.1103/PhysRevC.83.065202}{{\em Phys. Rev.} {\bf
  C83} (2011) 065202} [\href{http://arXiv.org/abs/1011.1988}{{\tt
  arXiv:1011.1988 [hep-ph]}}].

\bibitem{Accardi:2012qut}
A.~Accardi {\em et.~al.}, {\it {Electron Ion Collider: The Next QCD Frontier}},
   \href{http://dx.doi.org/10.1140/epja/i2016-16268-9}{{\em Eur. Phys. J.} {\bf
  A52} (2016)~no.~9 268} [\href{http://arXiv.org/abs/1212.1701}{{\tt
  arXiv:1212.1701 [nucl-ex]}}].

\bibitem{Aschenauer:2017jsk}
E.~C. Aschenauer, S.~Fazio, J.~H. Lee, H.~Mantysaari, B.~S. Page, B.~Schenke,
  T.~Ullrich, R.~Venugopalan and P.~Zurita, {\it {The electron–ion collider:
  assessing the energy dependence of key measurements}},
  \href{http://dx.doi.org/10.1088/1361-6633/aaf216}{{\em Rept. Prog. Phys.}
  {\bf 82} (2019)~no.~2 024301} [\href{http://arXiv.org/abs/1708.01527}{{\tt
  arXiv:1708.01527 [nucl-ex]}}].

\bibitem{AbelleiraFernandez:2012cc}
{\bf LHeC Study Group} collaboration, J.~L. Abelleira~Fernandez {\em et.~al.},
  {\it {A Large Hadron Electron Collider at CERN: Report on the Physics and
  Design Concepts for Machine and Detector}},
  \href{http://dx.doi.org/10.1088/0954-3899/39/7/075001}{{\em J. Phys.} {\bf
  G39} (2012) 075001} [\href{http://arXiv.org/abs/1206.2913}{{\tt
  arXiv:1206.2913 [physics.acc-ph]}}].

\bibitem{Chen:2018wyz}
X.~Chen, {\it {A Plan for Electron Ion Collider in China}},
  \href{http://dx.doi.org/10.22323/1.316.0170}{{\em PoS} {\bf DIS2018} (2018)
  170} [\href{http://arXiv.org/abs/1809.00448}{{\tt arXiv:1809.00448
  [nucl-ex]}}].

\bibitem{Melosh:1974cu}
H.~J. Melosh, {\it {Quarks: Currents and constituents}},
  \href{http://dx.doi.org/10.1103/PhysRevD.9.1095}{{\em Phys. Rev.} {\bf D9}
  (1974) 1095}.

\bibitem{Hufner:2000jb}
J.~Hufner, {\relax Yu}.~P. Ivanov, B.~Z. Kopeliovich and A.~V. Tarasov, {\it
  {Photoproduction of charmonia and total charmonium proton cross-sections}},
  \href{http://dx.doi.org/10.1103/PhysRevD.62.094022}{{\em Phys. Rev.} {\bf
  D62} (2000) 094022} [\href{http://arXiv.org/abs/hep-ph/0007111}{{\tt
  arXiv:hep-ph/0007111 [hep-ph]}}].

\bibitem{Kowalski:2006hc}
H.~Kowalski, L.~Motyka and G.~Watt, {\it {Exclusive diffractive processes at
  HERA within the dipole picture}},
  \href{http://dx.doi.org/10.1103/PhysRevD.74.074016}{{\em Phys. Rev.} {\bf
  D74} (2006) 074016} [\href{http://arXiv.org/abs/hep-ph/0606272}{{\tt
  arXiv:hep-ph/0606272 [hep-ph]}}].

\bibitem{Hatta:2017cte}
Y.~Hatta, B.-W. Xiao and F.~Yuan, {\it {Gluon Tomography from Deeply Virtual
  Compton Scattering at Small-$x$}},
  \href{http://dx.doi.org/10.1103/PhysRevD.95.114026}{{\em Phys. Rev.} {\bf
  D95} (2017)~no.~11 114026} [\href{http://arXiv.org/abs/1703.02085}{{\tt
  arXiv:1703.02085 [hep-ph]}}].

\bibitem{Shuvaev:1999ce}
A.~G. Shuvaev, K.~J. Golec-Biernat, A.~D. Martin and M.~G. Ryskin, {\it {Off
  diagonal distributions fixed by diagonal partons at small x and xi}},
  \href{http://dx.doi.org/10.1103/PhysRevD.60.014015}{{\em Phys. Rev.} {\bf
  D60} (1999) 014015} [\href{http://arXiv.org/abs/hep-ph/9902410}{{\tt
  arXiv:hep-ph/9902410 [hep-ph]}}].

\bibitem{Mantysaari:2016ykx}
H.~Mäntysaari and B.~Schenke, {\it {Evidence of strong proton shape
  fluctuations from incoherent diffraction}},
  \href{http://dx.doi.org/10.1103/PhysRevLett.117.052301}{{\em Phys. Rev.
  Lett.} {\bf 117} (2016)~no.~5 052301}
  [\href{http://arXiv.org/abs/1603.04349}{{\tt arXiv:1603.04349 [hep-ph]}}].

\bibitem{Cepila:2016uku}
J.~Cepila, J.~G. Contreras and J.~D. Tapia~Takaki, {\it {Energy dependence of
  dissociative $\mathrm{J/}\psi$ photoproduction as a signature of gluon
  saturation at the LHC}},
  \href{http://dx.doi.org/10.1016/j.physletb.2016.12.063}{{\em Phys. Lett.}
  {\bf B766} (2017) 186} [\href{http://arXiv.org/abs/1608.07559}{{\tt
  arXiv:1608.07559 [hep-ph]}}].

\bibitem{Mantysaari:2016jaz}
H.~Mäntysaari and B.~Schenke, {\it {Revealing proton shape fluctuations with
  incoherent diffraction at high energy}},
  \href{http://dx.doi.org/10.1103/PhysRevD.94.034042}{{\em Phys. Rev.} {\bf
  D94} (2016)~no.~3 034042} [\href{http://arXiv.org/abs/1607.01711}{{\tt
  arXiv:1607.01711 [hep-ph]}}].

\bibitem{Traini:2018hxd}
M.~C. Traini and J.-P. Blaizot, {\it {Diffractive incoherent vector meson
  production off protons: a quark model approach to gluon fluctuation
  effects}},  \href{http://dx.doi.org/10.1140/epjc/s10052-019-6826-0}{{\em Eur.
  Phys. J.} {\bf C79} (2019)~no.~4 327}
  [\href{http://arXiv.org/abs/1804.06110}{{\tt arXiv:1804.06110 [hep-ph]}}].

\bibitem{Mantysaari:2020axf}
H.~Mäntysaari, {\it {Review of proton and nuclear shape fluctuations at high
  energy}},  \href{http://arXiv.org/abs/2001.10705}{{\tt arXiv:2001.10705
  [hep-ph]}}.

\bibitem{Brodsky:1997de}
S.~J. Brodsky, H.-C. Pauli and S.~S. Pinsky, {\it {Quantum chromodynamics and
  other field theories on the light cone}},
  \href{http://dx.doi.org/10.1016/S0370-1573(97)00089-6}{{\em Phys. Rept.} {\bf
  301} (1998) 299} [\href{http://arXiv.org/abs/hep-ph/9705477}{{\tt
  arXiv:hep-ph/9705477 [hep-ph]}}].

\bibitem{Kovchegov:2012mbw}
Y.~V. Kovchegov and E.~Levin, {\it {Quantum chromodynamics at high energy}},
  \href{http://dx.doi.org/10.1017/CBO9781139022187}{{\em Camb. Monogr. Part.
  Phys. Nucl. Phys. Cosmol.} {\bf 33} (2012) 1}.

\bibitem{Lepage:1980fj}
G.~P. Lepage and S.~J. Brodsky, {\it {Exclusive Processes in Perturbative
  Quantum Chromodynamics}},
  \href{http://dx.doi.org/10.1103/PhysRevD.22.2157}{{\em Phys. Rev.} {\bf D22}
  (1980) 2157}.

\bibitem{Dosch:1996ss}
H.~G. Dosch, T.~Gousset, G.~Kulzinger and H.~J. Pirner, {\it {Vector meson
  leptoproduction and nonperturbative gluon fluctuations in QCD}},
  \href{http://dx.doi.org/10.1103/PhysRevD.55.2602}{{\em Phys. Rev.} {\bf D55}
  (1997) 2602} [\href{http://arXiv.org/abs/hep-ph/9608203}{{\tt
  arXiv:hep-ph/9608203 [hep-ph]}}].

\bibitem{Soper:1972xc}
D.~E. Soper, {\it {Infinite-momentum helicity states}},
  \href{http://dx.doi.org/10.1103/PhysRevD.5.1956}{{\em Phys. Rev.} {\bf D5}
  (1972) 1956}.

\bibitem{JalilianMarian:1996xn}
J.~Jalilian-Marian, A.~Kovner, L.~D. McLerran and H.~Weigert, {\it {The
  Intrinsic glue distribution at very small $x$}},
  \href{http://dx.doi.org/10.1103/PhysRevD.55.5414}{{\em Phys. Rev.} {\bf D55}
  (1997) 5414} [\href{http://arXiv.org/abs/hep-ph/9606337}{{\tt
  arXiv:hep-ph/9606337 [hep-ph]}}].

\bibitem{JalilianMarian:1997jx}
J.~Jalilian-Marian, A.~Kovner, A.~Leonidov and H.~Weigert, {\it {The BFKL
  equation from the Wilson renormalization group}},
  \href{http://dx.doi.org/10.1016/S0550-3213(97)00440-9}{{\em Nucl. Phys.} {\bf
  B504} (1997) 415} [\href{http://arXiv.org/abs/hep-ph/9701284}{{\tt
  arXiv:hep-ph/9701284 [hep-ph]}}].

\bibitem{JalilianMarian:1997gr}
J.~Jalilian-Marian, A.~Kovner, A.~Leonidov and H.~Weigert, {\it {The Wilson
  renormalization group for low x physics: Towards the high density regime}},
  \href{http://dx.doi.org/10.1103/PhysRevD.59.014014}{{\em Phys. Rev.} {\bf
  D59} (1998) 014014} [\href{http://arXiv.org/abs/hep-ph/9706377}{{\tt
  arXiv:hep-ph/9706377 [hep-ph]}}].

\bibitem{Iancu:2001md}
E.~Iancu and L.~D. McLerran, {\it {Saturation and universality in QCD at small
  $x$}},  \href{http://dx.doi.org/10.1016/S0370-2693(01)00526-3}{{\em Phys.
  Lett.} {\bf B510} (2001) 145}
  [\href{http://arXiv.org/abs/hep-ph/0103032}{{\tt arXiv:hep-ph/0103032
  [hep-ph]}}].

\bibitem{Ferreiro:2001qy}
E.~Ferreiro, E.~Iancu, A.~Leonidov and L.~McLerran, {\it {Nonlinear gluon
  evolution in the color glass condensate. 2.}},
  \href{http://dx.doi.org/10.1016/S0375-9474(01)01329-X}{{\em Nucl. Phys.} {\bf
  A703} (2002) 489} [\href{http://arXiv.org/abs/hep-ph/0109115}{{\tt
  arXiv:hep-ph/0109115 [hep-ph]}}].

\bibitem{Iancu:2001ad}
E.~Iancu, A.~Leonidov and L.~D. McLerran, {\it {The Renormalization group
  equation for the color glass condensate}},
  \href{http://dx.doi.org/10.1016/S0370-2693(01)00524-X}{{\em Phys. Lett.} {\bf
  B510} (2001) 133} [\href{http://arXiv.org/abs/hep-ph/0102009}{{\tt
  arXiv:hep-ph/0102009 [hep-ph]}}].

\bibitem{Iancu:2000hn}
E.~Iancu, A.~Leonidov and L.~D. McLerran, {\it {Nonlinear gluon evolution in
  the color glass condensate. 1.}},
  \href{http://dx.doi.org/10.1016/S0375-9474(01)00642-X}{{\em Nucl. Phys.} {\bf
  A692} (2001) 583} [\href{http://arXiv.org/abs/hep-ph/0011241}{{\tt
  arXiv:hep-ph/0011241 [hep-ph]}}].

\bibitem{Balitsky:1995ub}
I.~Balitsky, {\it {Operator expansion for high-energy scattering}},
  \href{http://dx.doi.org/10.1016/0550-3213(95)00638-9}{{\em Nucl. Phys.} {\bf
  B463} (1996) 99} [\href{http://arXiv.org/abs/hep-ph/9509348}{{\tt
  arXiv:hep-ph/9509348 [hep-ph]}}].

\bibitem{Kovchegov:1999yj}
Y.~V. Kovchegov, {\it {Small $x$ $F_2$ structure function of a nucleus
  including multiple pomeron exchanges}},
  \href{http://dx.doi.org/10.1103/PhysRevD.60.034008}{{\em Phys. Rev.} {\bf
  D60} (1999) 034008} [\href{http://arXiv.org/abs/hep-ph/9901281}{{\tt
  arXiv:hep-ph/9901281 [hep-ph]}}].

\bibitem{Albacete:2010sy}
J.~L. Albacete, N.~Armesto, J.~G. Milhano, P.~Quiroga-Arias and C.~A. Salgado,
  {\it {AAMQS: A non-linear QCD analysis of new HERA data at small-$x$
  including heavy quarks}},
  \href{http://dx.doi.org/10.1140/epjc/s10052-011-1705-3}{{\em Eur. Phys. J.}
  {\bf C71} (2011) 1705} [\href{http://arXiv.org/abs/1012.4408}{{\tt
  arXiv:1012.4408 [hep-ph]}}].

\bibitem{Lappi:2012nh}
T.~Lappi and H.~Mantysaari, {\it {Forward dihadron correlations in
  deuteron-gold collisions with the Gaussian approximation of JIMWLK}},
  \href{http://dx.doi.org/10.1016/j.nuclphysa.2013.03.017}{{\em Nucl. Phys.}
  {\bf A908} (2013) 51} [\href{http://arXiv.org/abs/1209.2853}{{\tt
  arXiv:1209.2853 [hep-ph]}}].

\bibitem{GolecBiernat:2003ym}
K.~J. Golec-Biernat and A.~M. Stasto, {\it {On solutions of the
  Balitsky-Kovchegov equation with impact parameter}},
  \href{http://dx.doi.org/10.1016/j.nuclphysb.2003.07.011}{{\em Nucl. Phys.}
  {\bf B668} (2003) 345} [\href{http://arXiv.org/abs/hep-ph/0306279}{{\tt
  arXiv:hep-ph/0306279 [hep-ph]}}].

\bibitem{Berger:2011ew}
J.~Berger and A.~M. Stasto, {\it {Small x nonlinear evolution with impact
  parameter and the structure function data}},
  \href{http://dx.doi.org/10.1103/PhysRevD.84.094022}{{\em Phys. Rev.} {\bf
  D84} (2011) 094022} [\href{http://arXiv.org/abs/1106.5740}{{\tt
  arXiv:1106.5740 [hep-ph]}}].

\bibitem{Berger:2012wx}
J.~Berger and A.~M. Stasto, {\it {Exclusive vector meson production and small-x
  evolution}},  \href{http://dx.doi.org/10.1007/JHEP01(2013)001}{{\em JHEP}
  {\bf 01} (2013) 001} [\href{http://arXiv.org/abs/1205.2037}{{\tt
  arXiv:1205.2037 [hep-ph]}}].

\bibitem{Mantysaari:2018zdd}
H.~Mäntysaari and B.~Schenke, {\it {Confronting impact parameter dependent
  JIMWLK evolution with HERA data}},
  \href{http://dx.doi.org/10.1103/PhysRevD.98.034013}{{\em Phys. Rev.} {\bf
  D98} (2018)~no.~3 034013} [\href{http://arXiv.org/abs/1806.06783}{{\tt
  arXiv:1806.06783 [hep-ph]}}].

\bibitem{Bendova:2019psy}
D.~Bendova, J.~Cepila, J.~G. Contreras and M.~Matas, {\it {Solution to the
  Balitsky-Kovchegov equation with the collinearly improved kernel including
  impact-parameter dependence}},
  \href{http://dx.doi.org/10.1103/PhysRevD.100.054015}{{\em Phys. Rev.} {\bf
  D100} (2019)~no.~5 054015} [\href{http://arXiv.org/abs/1907.12123}{{\tt
  arXiv:1907.12123 [hep-ph]}}].

\bibitem{Schlichting:2014ipa}
S.~Schlichting and B.~Schenke, {\it {The shape of the proton at high
  energies}},  \href{http://dx.doi.org/10.1016/j.physletb.2014.10.068}{{\em
  Phys. Lett.} {\bf B739} (2014) 313}
  [\href{http://arXiv.org/abs/1407.8458}{{\tt arXiv:1407.8458 [hep-ph]}}].

\bibitem{Kowalski:2003hm}
H.~Kowalski and D.~Teaney, {\it {An Impact parameter dipole saturation model}},
   \href{http://dx.doi.org/10.1103/PhysRevD.68.114005}{{\em Phys. Rev.} {\bf
  D68} (2003) 114005} [\href{http://arXiv.org/abs/hep-ph/0304189}{{\tt
  arXiv:hep-ph/0304189 [hep-ph]}}].

\bibitem{Gribov:1972ri}
V.~N. Gribov and L.~N. Lipatov, {\it {Deep inelastic e p scattering in
  perturbation theory}},  {\em Sov. J. Nucl. Phys.} {\bf 15} (1972) 438.
\newblock [Yad. Fiz.15,781(1972)].

\bibitem{Gribov:1972rt}
V.~N. Gribov and L.~N. Lipatov, {\it {$e^+ e^-$ pair annihilation and deep
  inelastic e p scattering in perturbation theory}},  {\em Sov. J. Nucl. Phys.}
  {\bf 15} (1972) 675.
\newblock [Yad. Fiz.15,1218(1972)].

\bibitem{Altarelli:1977zs}
G.~Altarelli and G.~Parisi, {\it {Asymptotic Freedom in Parton Language}},
  \href{http://dx.doi.org/10.1016/0550-3213(77)90384-4}{{\em Nucl. Phys.} {\bf
  B126} (1977) 298}.

\bibitem{Dokshitzer:1977sg}
Y.~L. Dokshitzer, {\it {Calculation of the Structure Functions for Deep
  Inelastic Scattering and e+ e- Annihilation by Perturbation Theory in Quantum
  Chromodynamics.}},  {\em Sov. Phys. JETP} {\bf 46} (1977) 641.
\newblock [Zh. Eksp. Teor. Fiz.73,1216(1977)].

\bibitem{Aaron:2009aa}
{\bf H1, ZEUS} collaborations, F.~D. Aaron {\em et.~al.}, {\it {Combined
  Measurement and QCD Analysis of the Inclusive $e^\pm p$ Scattering Cross
  Sections at HERA}},  \href{http://dx.doi.org/10.1007/JHEP01(2010)109}{{\em
  JHEP} {\bf 01} (2010) 109} [\href{http://arXiv.org/abs/0911.0884}{{\tt
  arXiv:0911.0884 [hep-ex]}}].

\bibitem{Abramowicz:2015mha}
{\bf H1, ZEUS} collaborations, H.~Abramowicz {\em et.~al.}, {\it {Combination of
  measurements of inclusive deep inelastic ${e^{\pm }p}$ scattering cross
  sections and QCD analysis of HERA data}},
  \href{http://dx.doi.org/10.1140/epjc/s10052-015-3710-4}{{\em Eur. Phys. J.}
  {\bf C75} (2015)~no.~12 580} [\href{http://arXiv.org/abs/1506.06042}{{\tt
  arXiv:1506.06042 [hep-ex]}}].

\bibitem{H1:2018flt}
{\bf H1, ZEUS} collaborations, H.~Abramowicz {\em et.~al.}, {\it {Combination
  and QCD analysis of charm and beauty production cross-section measurements in
  deep inelastic $ep$ scattering at HERA}},
  \href{http://dx.doi.org/10.1140/epjc/s10052-018-5848-3}{{\em Eur. Phys. J.}
  {\bf C78} (2018)~no.~6 473} [\href{http://arXiv.org/abs/1804.01019}{{\tt
  arXiv:1804.01019 [hep-ex]}}].

\bibitem{Abramowicz:1900rp}
{\bf H1, ZEUS} collaborations, H.~Abramowicz {\em et.~al.}, {\it {Combination
  and QCD Analysis of Charm Production Cross Section Measurements in
  Deep-Inelastic ep Scattering at HERA}},
  \href{http://dx.doi.org/10.1140/epjc/s10052-013-2311-3}{{\em Eur. Phys. J.}
  {\bf C73} (2013)~no.~2 2311} [\href{http://arXiv.org/abs/1211.1182}{{\tt
  arXiv:1211.1182 [hep-ex]}}].

\bibitem{Mantysaari:2018nng}
H.~Mäntysaari and P.~Zurita, {\it {In depth analysis of the combined HERA data
  in the dipole models with and without saturation}},
  \href{http://dx.doi.org/10.1103/PhysRevD.98.036002}{{\em Phys. Rev.} {\bf
  D98} (2018) 036002} [\href{http://arXiv.org/abs/1804.05311}{{\tt
  arXiv:1804.05311 [hep-ph]}}].

\bibitem{Rezaeian:2012ji}
A.~H. Rezaeian, M.~Siddikov, M.~Van~de Klundert and R.~Venugopalan, {\it
  {Analysis of combined HERA data in the Impact-Parameter dependent Saturation
  model}},  \href{http://dx.doi.org/10.1103/PhysRevD.87.034002}{{\em Phys.
  Rev.} {\bf D87} (2013)~no.~3 034002}
  [\href{http://arXiv.org/abs/1212.2974}{{\tt arXiv:1212.2974 [hep-ph]}}].

\bibitem{Bodwin:1994jh}
G.~T. Bodwin, E.~Braaten and G.~P. Lepage, {\it {Rigorous QCD analysis of
  inclusive annihilation and production of heavy quarkonium}},
  \href{http://dx.doi.org/10.1103/PhysRevD.55.5853,
  10.1103/PhysRevD.51.1125}{{\em Phys. Rev.} {\bf D51} (1995) 1125}
  [\href{http://arXiv.org/abs/hep-ph/9407339}{{\tt arXiv:hep-ph/9407339
  [hep-ph]}}].
\newblock [Erratum: Phys. Rev.D55,5853(1997)].

\bibitem{Braguta:2006wr}
V.~V. Braguta, A.~K. Likhoded and A.~V. Luchinsky, {\it {The Study of leading
  twist light cone wave function of $eta_c$ meson}},
  \href{http://dx.doi.org/10.1016/j.physletb.2007.01.014}{{\em Phys. Lett.}
  {\bf B646} (2007) 80} [\href{http://arXiv.org/abs/hep-ph/0611021}{{\tt
  arXiv:hep-ph/0611021 [hep-ph]}}].

\bibitem{Braguta:2007fh}
V.~V. Braguta, {\it {The study of leading twist light cone wave functions of
  $\mathrm{J}/\psi$ meson}},
  \href{http://dx.doi.org/10.1103/PhysRevD.75.094016}{{\em Phys. Rev.} {\bf
  D75} (2007) 094016} [\href{http://arXiv.org/abs/hep-ph/0701234}{{\tt
  arXiv:hep-ph/0701234 [HEP-PH]}}].

\bibitem{Bodwin:2007fz}
G.~T. Bodwin, H.~S. Chung, D.~Kang, J.~Lee and C.~Yu, {\it {Improved
  determination of color-singlet nonrelativistic QCD matrix elements for S-wave
  charmonium}},  \href{http://dx.doi.org/10.1103/PhysRevD.77.094017}{{\em Phys.
  Rev.} {\bf D77} (2008) 094017} [\href{http://arXiv.org/abs/0710.0994}{{\tt
  arXiv:0710.0994 [hep-ph]}}].

\bibitem{Bodwin:2006dn}
G.~T. Bodwin, D.~Kang and J.~Lee, {\it {Potential-model calculation of an
  order-$v^2$ NRQCD matrix element}},
  \href{http://dx.doi.org/10.1103/PhysRevD.74.014014}{{\em Phys. Rev.} {\bf
  D74} (2006) 014014} [\href{http://arXiv.org/abs/hep-ph/0603186}{{\tt
  arXiv:hep-ph/0603186 [hep-ph]}}].

\bibitem{Escobedo:2019bxn}
M.~Escobedo and T.~Lappi, {\it {Dipole picture and the nonrelativistic
  expansion}},  \href{http://dx.doi.org/10.1103/PhysRevD.101.034030}{{\em Phys.
  Rev.} {\bf D101} (2020)~no.~3 034030}
  [\href{http://arXiv.org/abs/1911.01136}{{\tt arXiv:1911.01136 [hep-ph]}}].

\bibitem{Flett:2019pux}
C.~A. Flett, S.~P. Jones, A.~D. Martin, M.~G. Ryskin and T.~Teubner, {\it {How
  to include exclusive $J/\psi$ production data in global PDF analyses}},
  \href{http://dx.doi.org/10.1103/PhysRevD.101.094011}{{\em Phys. Rev.} {\bf
  D101} (2020)~no.~9 094011} [\href{http://arXiv.org/abs/1908.08398}{{\tt
  arXiv:1908.08398 [hep-ph]}}].

\bibitem{Bodwin:2006dm}
G.~T. Bodwin, D.~Kang and J.~Lee, {\it {Reconciling the light-cone and NRQCD
  approaches to calculating $e^+ e^- \to \mathrm{J}/\psi + \eta_c$}},
  \href{http://dx.doi.org/10.1103/PhysRevD.74.114028}{{\em Phys. Rev.} {\bf
  D74} (2006) 114028} [\href{http://arXiv.org/abs/hep-ph/0603185}{{\tt
  arXiv:hep-ph/0603185 [hep-ph]}}].

\bibitem{Krassnigg:2001ka}
A.~Krassnigg and H.-C. Pauli, {\it {On helicity and spin on the light cone}},
  \href{http://dx.doi.org/10.1016/S0920-5632(02)01338-5}{{\em Nucl. Phys. Proc.
  Suppl.} {\bf 108} (2002) 251}
  [\href{http://arXiv.org/abs/hep-ph/0111260}{{\tt arXiv:hep-ph/0111260
  [hep-ph]}}].

\bibitem{Krelina:2019egg}
M.~Krelina, J.~Nemchik and R.~Pasechnik, {\it {$D$-wave effects in diffractive
  electroproduction of heavy quarkonia from the photon-like $V\rightarrow
  Q{\bar{Q}}$ transition}},
  \href{http://dx.doi.org/10.1140/epjc/s10052-020-7678-3}{{\em Eur. Phys. J.}
  {\bf C80} (2020)~no.~2 92} [\href{http://arXiv.org/abs/1909.12770}{{\tt
  arXiv:1909.12770 [hep-ph]}}].

\bibitem{Cepila:2019skb}
J.~Cepila, J.~Nemchik, M.~Krelina and R.~Pasechnik, {\it {Theoretical
  uncertainties in exclusive electroproduction of S-wave heavy quarkonia}},
  \href{http://dx.doi.org/10.1140/epjc/s10052-019-7016-9}{{\em Eur. Phys. J.}
  {\bf C79} (2019)~no.~6 495} [\href{http://arXiv.org/abs/1901.02664}{{\tt
  arXiv:1901.02664 [hep-ph]}}].

\bibitem{Krelina:2018hmt}
M.~Krelina, J.~Nemchik, R.~Pasechnik and J.~Cepila, {\it {Spin rotation effects
  in diffractive electroproduction of heavy quarkonia}},
  \href{http://dx.doi.org/10.1140/epjc/s10052-019-6666-y}{{\em Eur. Phys. J.}
  {\bf C79} (2019)~no.~2 154} [\href{http://arXiv.org/abs/1812.03001}{{\tt
  arXiv:1812.03001 [hep-ph]}}].

\bibitem{VanRoyen:1967nq}
R.~Van~Royen and V.~F. Weisskopf, {\it {Hadron Decay Processes and the Quark
  Model}},  \href{http://dx.doi.org/10.1007/BF02823542}{{\em Nuovo Cim.} {\bf
  A50} (1967) 617}.
\newblock [Erratum: Nuovo Cim.A51,583(1967)].

\bibitem{Tanabashi:2018oca}
{\bf Particle Data Group} collaboration, M.~Tanabashi {\em et.~al.}, {\it
  {Review of Particle Physics}},
  \href{http://dx.doi.org/10.1103/PhysRevD.98.030001}{{\em Phys. Rev.} {\bf
  D98} (2018)~no.~3 030001}.

\bibitem{Vary:2009gt}
J.~P. Vary, H.~Honkanen, J.~Li, P.~Maris, S.~J. Brodsky, A.~Harindranath, G.~F.
  de~Teramond, P.~Sternberg, E.~G. Ng and C.~Yang, {\it {Hamiltonian
  light-front field theory in a basis function approach}},
  \href{http://dx.doi.org/10.1103/PhysRevC.81.035205}{{\em Phys. Rev.} {\bf
  C81} (2010) 035205} [\href{http://arXiv.org/abs/0905.1411}{{\tt
  arXiv:0905.1411 [nucl-th]}}].

\bibitem{Honkanen:2010rc}
H.~Honkanen, P.~Maris, J.~P. Vary and S.~J. Brodsky, {\it {Electron in a
  transverse harmonic cavity}},
  \href{http://dx.doi.org/10.1103/PhysRevLett.106.061603}{{\em Phys. Rev.
  Lett.} {\bf 106} (2011) 061603} [\href{http://arXiv.org/abs/1008.0068}{{\tt
  arXiv:1008.0068 [hep-ph]}}].

\bibitem{Zhao:2014xaa}
X.~Zhao, H.~Honkanen, P.~Maris, J.~P. Vary and S.~J. Brodsky, {\it {Electron
  $g-2$ in Light-Front Quantization}},
  \href{http://dx.doi.org/10.1016/j.physletb.2014.08.020}{{\em Phys. Lett.}
  {\bf B737} (2014) 65} [\href{http://arXiv.org/abs/1402.4195}{{\tt
  arXiv:1402.4195 [nucl-th]}}].

\bibitem{Wiecki:2014ola}
P.~Wiecki, Y.~Li, X.~Zhao, P.~Maris and J.~P. Vary, {\it {Basis Light-Front
  Quantization Approach to Positronium}},
  \href{http://dx.doi.org/10.1103/PhysRevD.91.105009}{{\em Phys. Rev.} {\bf
  D91} (2015)~no.~10 105009} [\href{http://arXiv.org/abs/1404.6234}{{\tt
  arXiv:1404.6234 [nucl-th]}}].

\bibitem{Adhikari:2016idg}
L.~Adhikari, Y.~Li, X.~Zhao, P.~Maris, J.~P. Vary and A.~Abd El-Hady, {\it
  {Form Factors and Generalized Parton Distributions in Basis Light-Front
  Quantization}},  \href{http://dx.doi.org/10.1103/PhysRevC.93.055202}{{\em
  Phys. Rev.} {\bf C93} (2016)~no.~5 055202}
  [\href{http://arXiv.org/abs/1602.06027}{{\tt arXiv:1602.06027 [nucl-th]}}].

\bibitem{deTeramond:2008ht}
G.~F. de~Teramond and S.~J. Brodsky, {\it {Light-Front Holography: A First
  Approximation to QCD}},
  \href{http://dx.doi.org/10.1103/PhysRevLett.102.081601}{{\em Phys. Rev.
  Lett.} {\bf 102} (2009) 081601} [\href{http://arXiv.org/abs/0809.4899}{{\tt
  arXiv:0809.4899 [hep-ph]}}].

\bibitem{Brodsky:2014yha}
S.~J. Brodsky, G.~F. de~Teramond, H.~G. Dosch and J.~Erlich, {\it {Light-Front
  Holographic QCD and Emerging Confinement}},
  \href{http://dx.doi.org/10.1016/j.physrep.2015.05.001}{{\em Phys. Rept.} {\bf
  584} (2015) 1} [\href{http://arXiv.org/abs/1407.8131}{{\tt arXiv:1407.8131
  [hep-ph]}}].

\bibitem{Li:2015zda}
Y.~Li, P.~Maris, X.~Zhao and J.~P. Vary, {\it {Heavy Quarkonium in a
  Holographic Basis}},
  \href{http://dx.doi.org/10.1016/j.physletb.2016.04.065}{{\em Phys. Lett.}
  {\bf B758} (2016) 118} [\href{http://arXiv.org/abs/1509.07212}{{\tt
  arXiv:1509.07212 [hep-ph]}}].

\bibitem{Li:2017mlw}
Y.~Li, P.~Maris and J.~P. Vary, {\it {Quarkonium as a relativistic bound state
  on the light front}},
  \href{http://dx.doi.org/10.1103/PhysRevD.96.016022}{{\em Phys. Rev.} {\bf
  D96} (2017)~no.~1 016022} [\href{http://arXiv.org/abs/1704.06968}{{\tt
  arXiv:1704.06968 [hep-ph]}}].

\bibitem{Chen:2016dlk}
G.~Chen, Y.~Li, P.~Maris, K.~Tuchin and J.~P. Vary, {\it {Diffractive
  charmonium spectrum in high energy collisions in the basis light-front
  quantization approach}},
  \href{http://dx.doi.org/10.1016/j.physletb.2017.04.024}{{\em Phys. Lett.}
  {\bf B769} (2017) 477} [\href{http://arXiv.org/abs/1610.04945}{{\tt
  arXiv:1610.04945 [nucl-th]}}].

\bibitem{Chen:2018vdw}
G.~Chen, Y.~Li, K.~Tuchin and J.~P. Vary, {\it {Heavy quarkonia production at
  energies available at the CERN Large Hadron Collider and future electron-ion
  colliding facilities using basis light-front quantization wave functions}},
  \href{http://dx.doi.org/10.1103/PhysRevC.100.025208}{{\em Phys. Rev.} {\bf
  C100} (2019)~no.~2 025208} [\href{http://arXiv.org/abs/1811.01782}{{\tt
  arXiv:1811.01782 [nucl-th]}}].

\bibitem{Mantysaari:2017slo}
H.~Mäntysaari and R.~Venugopalan, {\it {Systematics of strong nuclear
  amplification of gluon saturation from exclusive vector meson production in
  high energy electron–nucleus collisions}},
  \href{http://dx.doi.org/10.1016/j.physletb.2018.04.044}{{\em Phys. Lett.}
  {\bf B781} (2018) 664} [\href{http://arXiv.org/abs/1712.02508}{{\tt
  arXiv:1712.02508 [nucl-th]}}].

\bibitem{Boussarie:2016bkq}
R.~Boussarie, A.~V. Grabovsky, D.~{\relax Yu}. Ivanov, L.~Szymanowski and
  S.~Wallon, {\it {Next-to-Leading Order Computation of Exclusive Diffractive
  Light Vector Meson Production in a Saturation Framework}},
  \href{http://dx.doi.org/10.1103/PhysRevLett.119.072002}{{\em Phys. Rev.
  Lett.} {\bf 119} (2017)~no.~7 072002}
  [\href{http://arXiv.org/abs/1612.08026}{{\tt arXiv:1612.08026 [hep-ph]}}].

\bibitem{Li:2018uif}
M.~Li, Y.~Li, P.~Maris and J.~P. Vary, {\it {Radiative transitions between
  $0^{-+}$ and $1^{--}$ heavy quarkonia on the light front}},
  \href{http://dx.doi.org/10.1103/PhysRevD.98.034024}{{\em Phys. Rev.} {\bf
  D98} (2018)~no.~3 034024} [\href{http://arXiv.org/abs/1803.11519}{{\tt
  arXiv:1803.11519 [hep-ph]}}].

\end{thebibliography}\endgroup

\end{document}